\documentclass[review]{elsarticle}

\usepackage{multirow}
\usepackage{multicol}
\usepackage{graphicx}
\usepackage{subfig}
\graphicspath{ {images/} }
\usepackage[top=2cm, bottom=2cm, left=2cm, right=2cm]{geometry}
\usepackage{algorithm}
\usepackage{algorithmicx}
\usepackage{algpseudocode}
\usepackage{amsmath}
\usepackage{color}

\journal{Journal of \LaTeX\ Templates}

\begin{document}

\begin{frontmatter}

\title{POBA-GA: Perturbation Optimized Black-Box Adversarial Attacks via Genetic Algorithm}



\author[mysecondaryaddress]{Jinyin~Chen\corref{mycorrespondingauthor}}
\cortext[mycorrespondingauthor]{Corresponding author}
\ead{chenjinyin@zjut.edu.cn}
\author[mysecondaryaddress]{Mengmeng~Su}
\ead{2111703406@zjut.edu.cn}
\author[mysecondaryaddress]{Shijing~Shen}
\ead{201407760128@zjut.edu.cn}
\author[mysecondaryaddress]{Hui~Xiong}
\ead{201403080329@zjut.edu.cn}
\author[mysecondaryaddress]{Haibin~Zheng}
\ead{2111703122@zjut.edu.cn}

\address[mysecondaryaddress]{Zhejiang University
of Technology, Hangzhou 310023, China}

\begin{abstract}
Most deep learning models are easily vulnerable to adversarial attacks. Various adversarial attacks are designed to evaluate the robustness of models and develop defense model. Currently, adversarial attacks are brought up to attack their own target model with their own evaluation metrics. And most of the black-box adversarial attack algorithms cannot achieve the expected success rate compared with white-box attacks. In this paper,comprehensive evaluation metrics are brought up for different adversarial attack methods. A novel perturbation optimized black-box adversarial attack based on genetic algorithm (POBA-GA) is proposed for achieving white-box comparable attack performances. Approximate optimal adversarial examples are evolved through evolutionary operations including initialization, selection, crossover and mutation. Fitness function is specifically designed to evaluate the example individual in both aspects of attack ability and perturbation control. Population diversity strategy is brought up in evolutionary process to promise the approximate optimal perturbations obtained. Comprehensive experiments are carried out to testify POBA-GA's performances. Both simulation and application results prove that our method is better than current state-of-art black-box attack methods in aspects of attack capability and perturbation control.
\end{abstract}


\end{frontmatter}


\section{Introduction}

Deep learning is the core of current machine learning and artificial intelligence~\cite{goodfellow2016deep}. Since it has powerful learning, feature extraction and modeling capabilities, it has been widely applied to challenging areas, such as social networks~\cite{deng2017deep}, medical image analysis~\cite{litjens2017survey, kooi2017large, havaei2017brain} and selective classification~\cite{geifman2017selective, de2018exploratory}. And in the area of computer vision, deep learning has become the main force for various applications such as self-driving cars~\cite{stilgoe2018machine}, image processing~\cite{szegedy2017processing, yosinski2014transferable}, target-driven visual navigation ~\cite{zhu2017target}and scene recognition~\cite{yuan2015scene}.

The latest research shows that although deep learning can extract complete image features and forecast or classify it perfectly, it can be easily fooled by adversarial examples into erroneous prediction outputs by adding small perturbations on the original image~\cite{liu2018security,ramanathan2017adversarial,bai2017alleviating,yin2018sparse,metzen2017universal}. The adversarial attack was first proposed by Szegedy et al.~\cite{szegedy2013intriguing} and attracted more attentions, becoming a new hot topic. As long as deep models are threatened by adversarial attack, lots of deep model based applications are unable to extend.{For instance, payment based on facial recognition cannot be trusted since an adversarial glass can help one imitate another person easily~\cite{sharif2016accessorize}, and auto-drive based on image recognition is quite dangerous if the road sign is a carefully designed adversarial example~\cite{eykholt2017robust}. In general, we can assess the robustness of the classifier by simulating the attacker's efforts to evade the classifier~\cite{xu2016automatically}, so it is also possible to increase the robustness of the model by defending adversarial attacks.} Lots of adversarial attack methods are brought up for understanding the attack and improving model's defensibility, such as Jacobian-based Saliency Map Attack (JSMA)~\cite{papernot2016limitations}, DeepFool~\cite{moosavi2016deepfool}, One-pixel Attack~\cite{Su2017One}, Limited Queries and Information Attack~\cite{ilyas2018black}.

Adversarial attacks can be roughly divided into three categories: gradient-based, score-based and transfer-based attacks~\cite{brendel2017decision}. Gradient-based and score-based attacks are often denoted as white-box and oracle attacks respectively. Most existing gradient-based white-box attacks rely on detailed model information, such as the Basic Iterative Model (BIM)~\cite{Kurakin2016Adversarial}, Houdini~\cite{Cisse2017Houdini}, DeepFool~\cite{moosavi2016deepfool}, which are taking advantage of gradient loss. There are also black-box attacks, which use transfer across models~\cite{papernot2017practical} or require access to all training data sets~\cite{Moosavidezfooli2017Universal}. Since most real-world systems do not publish the network's weight, architecture or training data sets, so white-box attacks and equivalent model attacks are not easy to implement in practice. There is still a flaw that if the attacker is capable of black-box attack without the internal configuration of the target model. That's the reason we develop a completely internal model information independent adversarial attack. It can not only fill the gap of evolutionary based adversarial attacks with minimal perturbation, but also help evaluating and improving the defensibility of current state-of-art deep models.

Excellent adversarial examples generally have the following two characteristics. First, it is very similar to the original image. There are slight perturbations barely distinguishable by the naked eye. Second, the adversarial example can lead the target model to a high confidence misclassification. Black-box attacks are conducted without any internal  model information (structure and parameter), but based on the information such as most probable class label or confidence. The target of black-box attack is to reduce the confidence of true label with limited perturbation. Therefore, in most cases black-box attack can be modeled as an optimization problem. Genetic algorithm is widely applied to various applications as a typical optimization tool, such as energy optimization~\cite{talha2017energy}, distribution network optimization~\cite{syahputra2017distribution}, ontology alignments optimization~\cite{gil2018optimizing} and web crawler~\cite{goyal2016genetic}, and all of them achieve good optimization performance. In this paper, a novel adversarial perturbation optimization attack based on genetic algorithm is proposed to implement black-box attack. Fitness function is constructed on the basis of classification confidence and perturbation size, and genetic operations are designed to promise approximate optimal adversarial example.

The current adversarial attacks generally use $L_{0}$, $L_{2}$ and $L_{\propto}$ as evaluation metrics for the perturbations. However, most attack algorithms use different perturbation evaluation metrics according to the characteristics of the algorithm~\cite{akhtar2018threat}. For example, L-BFGS~\cite{szegedy2013intriguing} and FGSM~\cite{goodfellow2014explaining} use $L_{\propto}$, JSMA~\cite{papernot2016limitations} and One-pixel~\cite{Su2017One} use $L_{0}$, DeepFool~\cite{moosavi2016deepfool} uses $L_{2}$. Therefore, we can only evaluate the quality of the algorithm from different perspectives, and cannot judge which perturbation and algorithm are better. Based on the sensitivity of the human visual system to perturbation, a new perturbation assessment method is put forward which can comprehensively evaluate the perturbation, making the perturbation more similar to the actual sensitivity.

The main contribution of our work can be concludes as:
\begin{itemize}
\item \textbf{\emph{Perturbation optimization.}} POBA-GA is a novel perturbation optimization method to generate black-box adversarial examples, which is capable of high successful attack rate against different deep learning models with controllable perturbations.
\item \textbf{\emph{White-box comparable black-box attack.}} POBA-GA can optimize the perturbations based on the confidence of the black-box output. In most cases, it can achieve high attack success rate comparable with white-box attack methods.
\item \textbf{\emph{New perturbations evaluation metric.}} A novel perturbation evaluation metrics is put forward. It comprehensively evaluates the perturbation and maps its size to different dimensions, which makes the evaluation result more realistic reflection of the perturbation.
\item \textbf{\emph{Improve defense capability through adversarial training.}} Adversarial examples generated by POBA-GA are adopted to train deep model to improve its defense capacity. Experiments prove that adversarial training based on examples from POBA-GA could defend model better than from other attack methods.
\end{itemize}
The rest of paper is organized as follows. The related works are discussed in Sec.~\ref{RW}. The main methods and strategies are introduced in Sec.~\ref{Approach}. Experiments and conclusions are shown in Sec.~\ref{Exp} and Sec.~\ref{Conclusion}.

\section{Related Works\label{RW}}

In this section, we introduce classic adversarial attack methods, genetic algorithms and perturbation evaluation metrics.

\subsection{Attacks methods}
Since Szegedy et al.~\cite{szegedy2013intriguing} proposed the concept of adversarial attack against deep model, a large number of adversarial attacks are put forward~\cite{goodfellow2014explaining, carlini2017towards, moosavi2016deepfool, moosavi2017universal, hayes2017machine}. Some researchers have launched attacks in applications, such as speech recognition systems~\cite{carlini2016hidden}, malware detectors~\cite{hu2017black, xu2016automatically}, and face recognition systems~\cite{sharif2017adversarial}. For example, Mengying Sun et al.~\cite{sun2018identify} use adversarial attacks against deep predictive models to identify susceptible locations in medical records. Tegjyot Singh Sethi et al.~\cite{sethi2018data} present an adversary's view point of a classification based system. Generally, attacks are classified into white-box attacks and black-box attacks based on whether they know the internal structure of the target model. It should be noted that, because there are too many adversarial attack literatures, this paper mainly introduces some algorithms of computer vision.

\subsubsection{White-box Adversarial Attack Models}
A white-box attack is a method of attacking a target model while it knows its internal structure. White-box attacks are the most common attack method, which can also be used to attack the equivalent model to achieve black-box attacks. At present, researchers have proposed a large number of white-box attacks~\cite{goodfellow2014explaining, Kurakin2016Adversarial, papernot2016limitations, moosavi2016deepfool, carlini2017towards}. For example, Bose et al. proposed adversarial attacks on face detectors using neural net based constrained optimization~\cite{bose2018adversarial}. Yu et al. proposed a fast adversarial attack example generation framework based on adversarial saliency prediction~\cite{yu2018asp}. Chen et al. proposed robust physical adversarial attack on faster R-CNN object detector~\cite{chen2018robust}. Ramanathan et al. proposed adversarial attacks on computer vision algorithms using natural perturbations~\cite{ramanathan2017adversarial}.

The perturbation optimization algorithm in this paper takes the adversarial examples generated by the white-box attack as a partial initial solution and realizes the perturbation optimization through the genetic algorithm. To ensure the diversity of the initial solution, we chose seven different attack methods to generate. The reasons for choosing this methods are described in detail in Sec.~\ref{Approach}. The attack method is described in detail below.

{\emph{\textbf{Fast Gradient Sign Model (FGSM)}}}~\cite{goodfellow2014explaining}
FGSM is one of the simplest and most widely used non-target counter attacks. FGSM uses backward propagation gradient from the target DNN to generate adversarial examples. Perturbation is evaluated by $\rho =\epsilon sign(\bigtriangledown J(\theta,I_{c},l))$[1], where $\bigtriangledown J$ denotes the gradient of the original image $I_{c}$ around the model parameters $\theta$. $sign(.)$ denotes the sign function, and $\epsilon$ is a small scalar value that limits the perturbation evaluation.

{\emph{\textbf{Basic Iterative Model (BIM)}}}~\cite{Kurakin2016Adversarial}
BIM, equivalent to Projected Gradient Descent, is a standard convex optimization method. It is an extension of the single-step method, which takes multiple small step iterations while adjusting the direction after each step. After a sufficient number of iterations, BIM can successfully generate an adversarial example classified into the target label.

{\emph{\textbf{Jacobian-based Saliency Map Attack (JSMA)}}}~\cite{papernot2016limitations}
JSMA describes the input-output relationship of the target DNN by constructing a Jacobian-based saliency map. It iteratively modifies the most important pixels based on saliency mapping during iteration to fool the network. At each iteration, JSMA recalculates the saliency map and uses the DNN derivative of the input image as a modify index of the adversarial attack. This greedy search process is repeated until the number of changing pixels reaches the threshold or the deception is successful.

{\emph{\textbf{DeepFool}}}~\cite{moosavi2016deepfool}
DeepFool is a simple but very effective non-targeted attack. In each iteration, it calculates the minimum distance $d(y_{1},y_{0})$ required for each label $y_{1}\ne y_{0}$ to reach the class boundary by approximating the model label with a linear label. $y_{1}$ represents the label of the adversarial example classified by deep model, and $y_0$ represents the true label of the adversarial example. Then make the appropriate steps in the direction of the nearest class. The image perturbation for each iteration are accumulated and the final perturbation is calculated once the output criteria changes.

{\emph{\textbf{Carlini and Wagner Attacks (C$\&$W)}}}~\cite{carlini2017towards}
Carlini and Wagner Attacks is one of the strongest attacks. Its attack essence is a kind of refined iterative gradient attack that uses the Adam optimizer. It uses the internal configuration of the target DNN to guide the attack, and uses $L_{2}$ specification to quantify the difference between the hostile and original image.

{\emph{\textbf{Gaussian Blur}}}~\cite{chen2009empirical}
Gaussian Blur is a kind of linear smoothing filter, which is suitable for eliminating Gaussian noise and is widely used in the noise reduction process of image processing. Generally speaking, Gaussian filtering is the process of weighted averaging of the entire image. The value of each pixel is obtained by weighted averaging of its own and other pixel values in the neighborhood. However, the Gaussian filter also causes the image to lose certain eigenvalues, making the CNN classification error.

{\emph{\textbf{Salt and Pepper Noise}}}~\cite{varatharajan2017adaptive}
Salt and pepper noise, also known as impulsive noise, randomly changes some pixel values. Salt and pepper noise appears on a binary image is to make some pixels white or black, black is pepper, and white is salt. And these salt and pepper noises have a certain probability of misclassifying the CNN label.

\subsubsection{Black-box Adversarial Attack Models}
The black-box can only access the input and output of the target model, but cannot access the internal configuration of the target model. In case of image label trained by CNN, we take the image as input and produces a confidence score for each label as an output~\cite{chen2017zoo}. In practical applications, most of the target models are black-box models, so black-box attacks have important research significance and have been studied by many researchers~\cite{smith2018understanding, dong2018boosting, milton2018evaluation}. For example, Milton et al.~\cite{milton2018evaluation} proposed the evaluation of momentum diverse input iterative fast gradient sign method (M-DI2-FGSM) to attack the black-box facial recognition system. Dong et al.~\cite{dong2018boosting} propose a broad class of momentum-based iterative algorithms to boost adversarial attacks. Brendel et al.~\cite{brendel2017decision} introduce the Boundary Attack, open new avenues to study the robustness of machine learning models and raise new questions regarding the safety of deployed machine learning systems. Andrew et al.~\cite{ilyas2018black} proposed the black-box adversarial attacks with limited information query. In the current black-box attack, ZOO and Boundary are the state-of-art attack methods, which aim to improve the attack rate.

\subsection{Genetic algorithm}
The existing attack method can be regarded as the solution of the optimization problem to some extent. For example, ZOO uses a zeroth order method to optimize black-box attack~\cite{chen2017zoo}. UPSET and ANGRI use the so-called UPSET network to optimize black-box attack~\cite{sarkar2017upset}. Houdini optimized for mean per-pixel or per-class accuracy instead of mIoU for better experimental results~\cite{Cisse2017Houdini}. Genetic algorithm has been widely used in various optimization problems and have achieved good results, for example, structures locations optimization~\cite{kalajacoptimization}, distribution network optimization~\cite{alencar2018optimal} and relay related optimization~\cite{souza2018optimized}. Therefore, we apply genetic algorithms to perturbation optimization to generate high-quality adversarial examples. Solutions are initialized as population. Fitness of each individual in population is calculated. Selection, crossover and mutation operators are carried out by certain probability to update whole population until termination condition is met. Approximate optimal solution will be found for the problem.

\subsection{Perturbation evaluation metrics}
Current adversarial attacks use $L_{0}$, $L_{2}$ and $L_{\propto}$ to evaluate the size of the perturbation. $L_{p}$ distance metric is used as a measure of similarity~\cite{hayes2017machine}, and it is used to evaluate the size of perturbation. The $L_{0}$ distance measurement indicates the number of pixels changed, and the $L_{2}$ represents the Euclidean distance between the two examples, and the $L_{\propto}$ represents the maximum perturbation between two pixels. Table~\ref{my-label: perturbation} list the perturbation metrics and other attributes of the current adversarial attack methods in computer vision. Due to the recent research on adversarial attacks is very popular, we are unable to list all the literature, so the table only lists the popular ones or representative of the popular direction of computer vision adversarial attack methods.

\begin{table*}[!t]
\centering
\caption{Attributes of the computer vision adversarial attack methods.}
\label{my-label: perturbation}
\begin{tabular}{l l c c c c}
\hline
\hline
Model                          & \multicolumn{1}{c}{Black/White-box} & Targeted/Non-targeted     & Specific/Universal        & Perturbation norm         & Learning                  \\ \hline
L-BFGS~\cite{szegedy2013intriguing}               & White-box          & Targeted           & Image specific        & $L_{\propto}$         & One shot            \\
FGSM~\cite{goodfellow2014explaining}             & White-box          & Targeted           & Image specific        & $L_{\propto}$         & One shot            \\
BIM \& ILCM~\cite{Kurakin2016Adversarial}        & White-box          & Non-targeted           & Image specific        & $L_{\propto}$         & Iterative            \\
JSMA~\cite{papernot2016limitations}              & White-box          & Targeted           & Image specific        & $L_{0}$         & Iterative            \\
One-pixel~\cite{Su2017One}                        & Black-box          & Non-targeted           & Image specific        & $L_{0}$         & Iterative            \\
C\&W~\cite{carlini2017towards}            & White-box          & Targeted           & Image specific        & $L_{0},L_{2},L_{\propto}$         & Iterative            \\
DeepFool~\cite{moosavi2016deepfool}               & White-box          & Non-targeted           & Image specific        & $L_{2},L_{\propto}$         & Iterative            \\
Uni.perturbations~\cite{moosavi2017universal}     & White-box          & Non-targeted           & Universal        & $L_{2},L_{\propto}$         & Iterative            \\
UPSET~\cite{sarkar2017upset}                      & Black-box          & Targeted           & Universal        & $L_{\propto}$         & Iterative            \\
ANGRI~\cite{sarkar2017upset}                      & Black-box          & Targeted           & Image specific        & $L_{\propto}$         & Iterative            \\
ZOO~\cite{chen2017zoo}                   & Black-box          & Non-targeted           & Image specific        & $L_{2}$         & Iterative            \\
Boundary~\cite{brendel2017decision}                   & Black-box          & Targeted           & Image specific        & $L_{2}$         & Iterative            \\
Limited~\cite{ilyas2018black}                   & Black-box          & Non-targeted           & Image specific        & $L_{\propto}$         & Iterative            \\
MI-FGSM~\cite{dong2018boosting}                   & Both         & Both           & Image specific        & $L_{2},L_{\propto}$         & Iterative            \\
AutoZOOM~\cite{Tu2018AutoZOOM}& Black-box         & Both           & Image specific        & $L_{2}$         & Iterative            \\

\hline

\hline
\end{tabular}
\end{table*}

\section{Model\label{Approach}}

\subsection{Problem definition}
For a given example $S$, Deep Neural Model (DNN) is applied to classify $S$ with an output label. Attack Method $AM$ is adopted to generate perturbation $A$ to add to $S$. Directly or optimally generate an adversarial example $AS$ to attack target model $TM$, so the target model outputs an error label. In the field of image recognition, $S$ represents the original image. The adversarial example $AS$ is generated and used to attack method $AM$, which is added by negligible perturbation $A$ .

\subsubsection{DEFINITION 1 (DNN based image label)}
Deep Neural Network (DNN) is trained by a large number of labeled images. For a given image $S$, DNN can output label $y_1$, represented as $TM(\Theta, S)=y_{1}$, where $\Theta$ represents parameters, $y_{1}$ is the output label of the highest confidence.

\subsubsection{DEFINITION 2 (Adversarial attack)}
Given a DNN for $S$, whose response output is $TM(\Theta, S)=y_{0}$, Attack Method $AM$ generates an adversarial image $AS$ to make $TM(\Theta, AS)=y_{1}$, and $y_{0}\ne y_{1}$, where $S$ and $AS$ are almost indistinguishable.

\captionsetup[figure]{singlelinecheck=off}
\begin{figure}[H]
\centering
\includegraphics[width=0.4\linewidth]{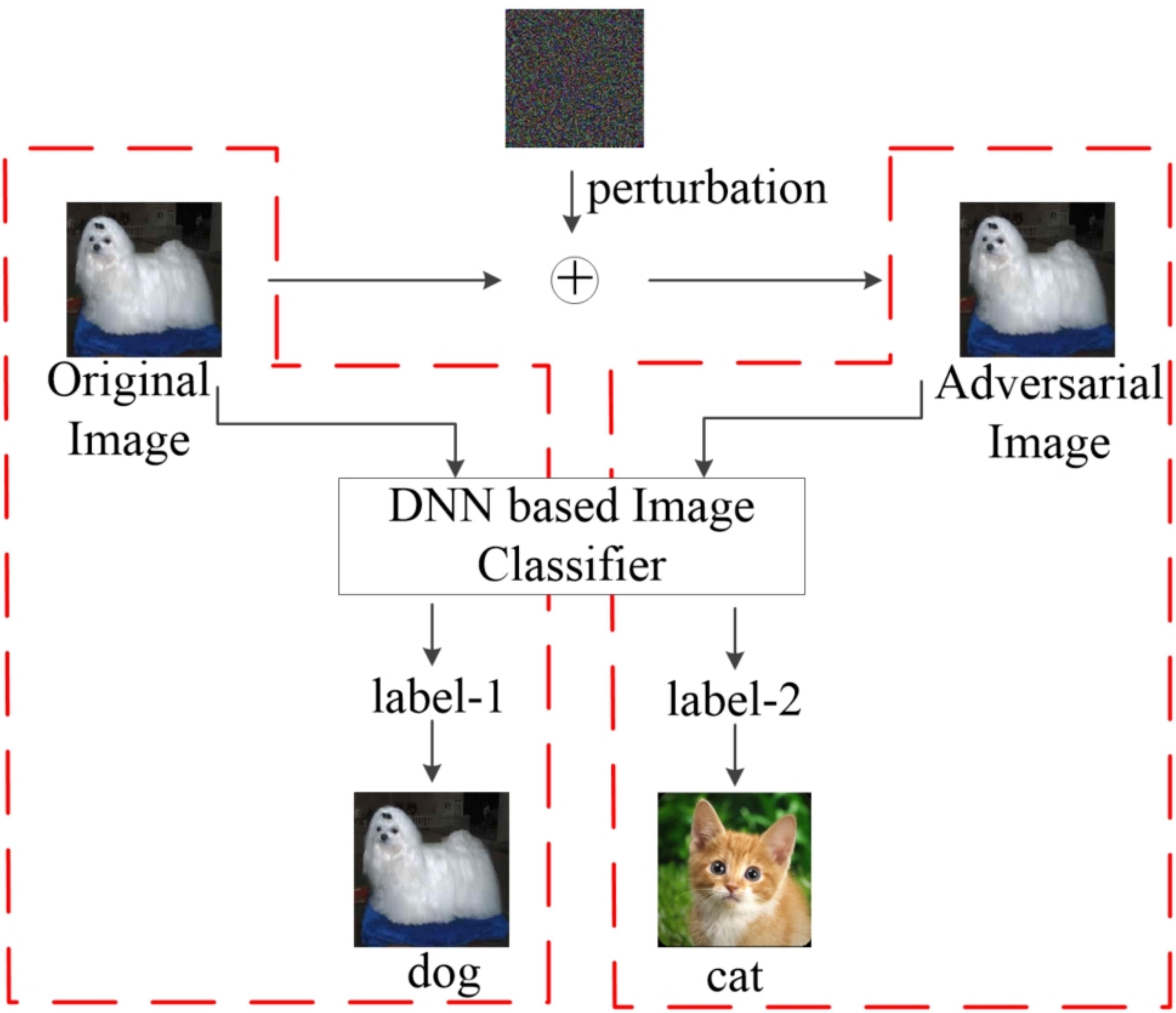}
\caption{A illustration of black-box attack. Black-box attacks can be viewed as choosing the best countermeasure example generated from a white-box attack or generating a better confrontation example than the current white-box attack.}
\label{fig:adversarial_attack}
\end{figure}

Adversarial attack is triggered by adversarial examples. Most adversarial examples are generated by adding perturbations into the original image. The perturbation quality will decide the attack capacity. An effective black-attack can be considered as generating a high quality perturbation without knowing the internal structure of the $TM$, causing the $TM$ to output an error class , which is illustrated in Figure \ref{fig:adversarial_attack}. And how to generated a black-box adversarial example can be assumed as an optimization problem.

\subsection{Framework}
\begin{figure*}[!t]
\centering
\includegraphics[width=1.0\linewidth]{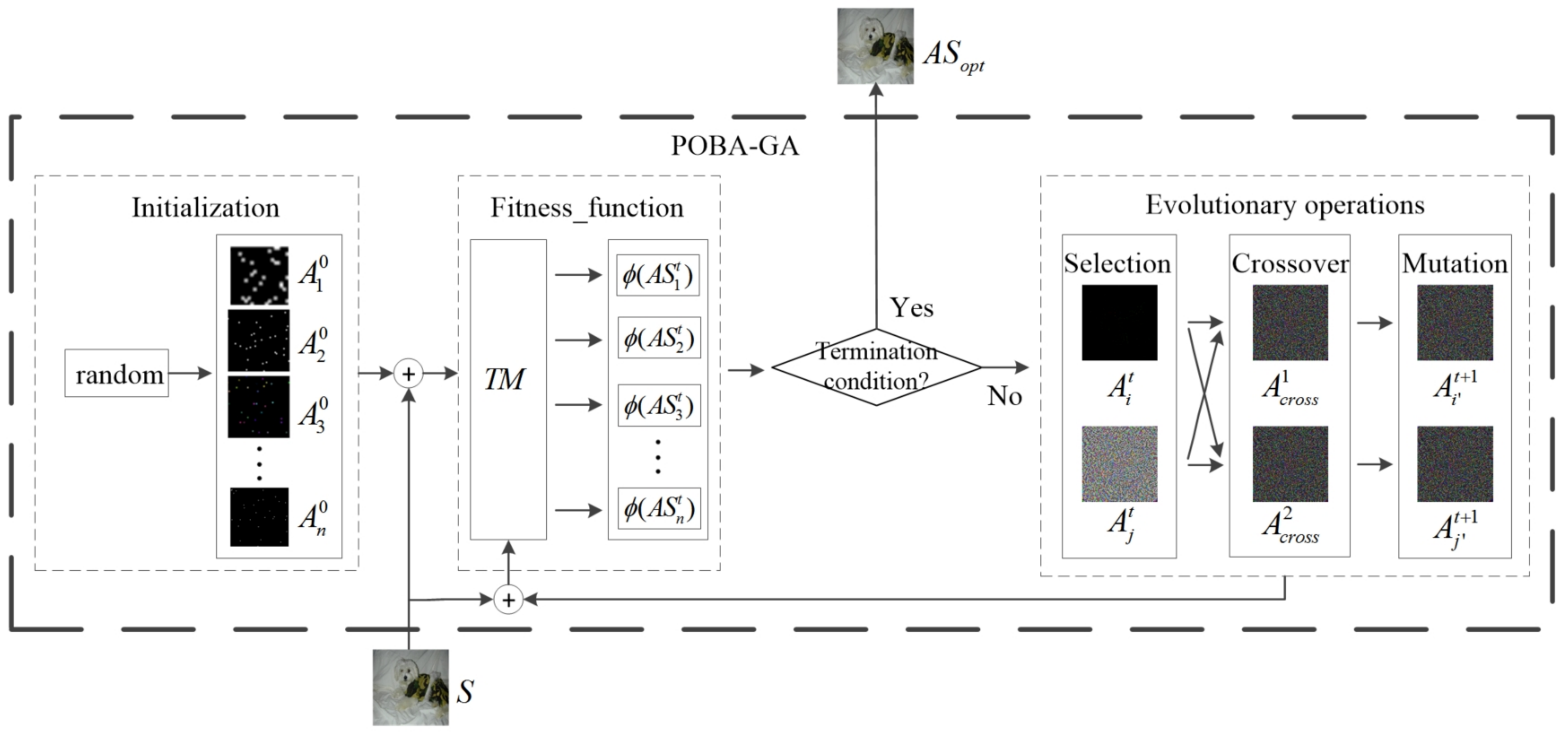}
\caption{Algorithm block diagram of POBA-GA. Black-box attack is translated into the problem of finding the best perturbation.
POBA-GA is established in stages:
1) Initialization. Generating a variety of different types of random perturbations.
2) Calculate fitness. Inputting the adversarial examples into the $TM$ to obtain their classification result. Then, the fitness is calculated based on the classification result and the perturbation size.
3)Determine if the termination condition is reached, go to the next step or output the best adversarial example.
4) Evolutionary operations. Use the roulette selection method to select the perturbations corresponding to the adversarial examples, and then obtain the next generation perturbation by crossing and mutating.
}
\label{fig:Fig1}
\end{figure*}

We propose a novel Perturbation Optimization black-box Attack based on Genetic Algorithm (POBA-GA), which takes a variety of different random perturbations as the initial examples, $\phi(.)$ is the fitness function to optimize the perturbation to obtain approximate optimal adversarial example $AS_{opt}$. The block diagram of POBA-GA is illustrated in Figure~\ref{fig:Fig1}. And the symbols used in the paper are listed in Table~\ref{my-tabel:symbols}.

Figure~\ref{fig:Fig1} demonstrates how a high quality adversarial example is generated through genetic algorithm. First, we generate different perturbations based on different noise point pixel thresholds, number of noise points, and noise point size. Then, add the perturbation into the original example $S$ to generate the responding initial adversarial example $AS^{t=0}$, where $t=0$ represents the first generation of population. Second, we define a fitness function $\phi(AS^{t}_{i})$ to evaluate the $t^{th}$ iteration example $AS^{t}_{i}\in AS^{t}$, $i\in[0,n]$. Third, we judge whether the termination condition has been reached. Fourth, we apply typical operators in genetic algorithm including selection, crossover and mutation to evolve new generation for perturbation optimization.

\subsection{Initialization}

Initialization is responsible for initial solution generation at the beginning. The quality of the initial solution directly affects the iterations. If the initial solution is similar to the approximate optimal solution, the algorithm will convergence quickly. Otherwise, the algorithm need more iterations to the convergence to approximate global optimal. In addition to consider the quality of each initial solution, the diversity of the initial solution is also crucial. It has been proved that diversity of initial population of genetic algorithm could promise approximate global optimal~\cite{konak2006multi}.

For a given example $S$, the optimization purpose is to evolve an approximate optimal adversarial example $AS_{opt}$. In random perturbation initialization, we use a search distribution of random Gaussian noise around the original image $S$~\cite{ilyas2018black}, which is described as $AS^{t=0}=S+\delta$, where $\delta\sim\aleph(\mu,\sigma^2)$ and $\mu$ represent mathematical expectation, $\sigma^2$ represent variance. . In order to increase the diversity of the initial perturbation, this paper generates different types of initial perturbation based on different variance, number of noise points, and noise point size.

\begin{table}[H]
\centering
\caption{The symbols used in the paper.}
\begin{tabular}{rl}
\hline %
$S$     &   the original example\\
$AS_{opt}$  &   the approximate optimal adversarial example of POBA-GA\\
$AS^{t}$    &   the collection of $t^{th}$ iteration adversarial examples\\
               & (when $t=0$, it represents the initial adversarial example)\\
$AS_{i}^{t}$    & the $t^{th}$ iteration of the $i^{th}$ adversarial example\\
$AS^{t}_{i(ab)}$    &   the pixels of $a^{th}$ row and $b^{th}$ column of $AS_{i}^{t}$\\
$A^{t}$    &   a collection of perturbation of the $t^{th}$ iteration\\
$A_{i}^{t}$    & the $t^{th}$ iteration of the $i^{th}$ perturbation\\
$TM$    &   the target method\\
$L_{0}, L_{2}, L_{\propto}$     &   zero/ two/ infinite norm\\
$\phi(AS^{t}_{i})$  &   the fitness function of $AS^{t}_{i}$\\
$P(AS^{t}_{i})$     &   the attack performance of $AS^{t}_{i}$ in $\phi(.)$\\
$Z(A^{t}_{i})$      &    the perturbation evaluation of $A^{t}_{i}$ in $\phi(.)$\\
$\alpha$    &   the perturbation ratio parameter in $\phi(.)$\\
$f(AS^{t}_{i})$     &   the selected probability of $AS^{t}_{i}$\\
$fr(AS^{t}_{i})$        &   the cumulative probability of $AS^{t}_{i}$\\
$p(y|AS^{t}_{i})$      &   the confidence of $AS^{t}_{i}$ labeled as $y$\\
$y_{0}$     &   the true label of the original example $S$\\
$y_{1},y_{2}$   &   the label for $S$ with first and second confidence\\
$y_{tar}$     &    the preset label for target attack\\
$B,C$     &   the two-dimensional matrix in crossover/ mutation\\
$ASR$       &   the Attack Success Rate\\
$P_{c}, P_{m}$     &   the crossover/ mutation probability\\
\hline %
\end{tabular}
\label{my-tabel:symbols}
\end{table}

\captionsetup[figure]{singlelinecheck=off}

{\subsection{Fitness function}

Fitness function is defined to evaluate the quality of examples in GA. The proper fitness function can directly affect the convergence speed of genetic algorithm and whether it can find the optimal solution.

Excellent adversarial examples generally have the following two characteristics. First, it is very similar to the original image. There are only slight perturbation that are barely distinguishable by the naked eye. Second, the adversarial example may be misclassified by the target model with a highly confidence. Therefore, the fitness function designed should be related to the confidence and perturbation size, Eq.~\ref{equ:Eq_2}.
\begin{equation}
\begin{array}{c}
\phi(AS^{t}_{i})= P(AS^{t}_{i})- \frac{\alpha}{\max Z(A^{0})} Z(A^{t}_{i})
\end{array}
\label{equ:Eq_2}
\end{equation}
where $\phi(AS^{t}_{i})$ represent the fitness function for example $AS^t_i$, $P(AS^{t}_{i})$ represent the attack performance of example $AS^{t}_{i}$ calculated by Eq.~\ref{equ:Eq_3}, $Z(A^{t}_{i})$ represents the size of adversarial examples perturbation can calculated by Eq.~\ref{equ:Eq_7}, and $A^{t}_{i}+S=AS^{t}_{i}$. $Z(A^{t}_{i})$ is a novel perturbation metric proposed in this paper, which will be explained and compared in Sec.~\ref{Evaluation Metrics}. $\alpha$ represents the proportional coefficient and is used to adjust the proportion of attack performance and perturbation size. For example, when $\alpha=0$, the fitness function only considers attack performance $P(AS^{t}_{i})$. When $\alpha$ is larger, the optimization process pays more attention to the size of the perturbation. The optimized adversarial example may have only general attack performance, but the perturbation is very small.

\begin{equation}
\begin{array}{c}
P(AS^{t}_{i})=\left\{
            \begin{array}{ll}
            p(y_{1}|AS^{t}_{i})-p(y_{0}|AS^{t}_{i}) & y_{1}\ne y_{0}\\
            p(y_{2}|AS^{t}_{i})-p(y_{0}|AS^{t}_{i}) & y_{1}=y_{0}
            \end{array}
\right.
\label{equ:Eq_3}
\end{array}
\end{equation}
where $y_{0}$ represents the true label of the adversarial image, and $y_{1}$, $y_{2}$ represent the label with highest and second highest confidence of the $TM$ output for $AS^t_i$, respectively. And $y_{1}$ also means the output label. $p(y|AS^{t}_{i})$ represents the confidence that the adversarial example $AS^t_i$ is labeled as $y$ by the $TM$. When the output label is different from $y_{0}$ of $AS^t_i$, the attack is successful. The $TM$ outputs the wrong label, and the attack performance is the confidence difference between the label $y_1$ and the true label $y_0$. The greater the attack capability, the stronger the attack capability. Otherwise, the output label is the same as the true label, which means the attack failed. Then we calculate the confidence difference between the highest label and the second highest label. The larger the difference, the more difficult it is to succeed.

In order to achieve the attack faster and reduce the initial number of queries to the DNN model, we will not consider the perturbation before the attack succeeds, i.e. let $\alpha=0$. Only when the attack is successful, we will consider both attack performance $P(AS^{t}_{i})$ and perturbation $Z(A^{t}_{i})$. The updated fitness function is as follows.

\begin{equation}
\begin{array}{c}
\phi(AS^{t}_{i})=\left\{
            \begin{array}{ll}
            p(y_{1}|AS^{t}_{i})-p(y_{0}|AS^{t}_{i})-\frac{\alpha}{\max Z(A^{t_0})} Z(A^{t}_{i}) & y_{1}\ne y_{0}\\
            p(y_{2}|AS^{t}_{i})-p(y_{0}|AS^{t}_{i}) & y_{1}=y_{0}
            \end{array}
\right.
\label{equ:Eq_fit}
\end{array}
\end{equation}
where $t_0$ is the number of iterations when the initial attack succeeds, and $\frac{\alpha}{\max Z(A^{t_0})}$ is used to control the perturbation to a certain range.

\subsection{Evolutionary operations}

\subsubsection{Selection operator}
Selection operator is adopted to choose two parent examples to produce two children by crossover and mutation operators. Generally, quality examples are more likely to be selected. This paper uses roulette wheel selection~\cite{lipowski2012roulette}. For a given example $AS^{t}_{i}$, its selection probability can be calculated according to Eq.~\ref{equ:Eq_f}. And the interval $(fr(AS^{t}_{i-1}),fr(AS^{t}_{i})]$ for each model is calculated according to Eq.~\ref{equ:Eq_fr}.

\begin{equation}
\begin{array}{c}
f(AS^{t}_{i})=\frac{\phi(AS^{t}_{i})}{\sum_{j=1}^{n}\phi(AS^{t}_{j})}
\label{equ:Eq_f}
\end{array}
\end{equation}

\begin{equation}
\begin{array}{c}
fr(AS^{k}_{i})={\sum_{j=1}^{i}f(AS^{t}_{j})}
\label{equ:Eq_fr}
\end{array}
\end{equation}

For better understanding, we give a simple illustration of Roulette wheel selection. Let $n=6$, table~\ref{my-label:Calculation results}} shows $\phi(AS^{t}_{i})$, $f(AS^{t}_{i})$ and $fr(AS^{t}_{i})$ of the adversarial examples. The table shows that the adversarial example $AS^t_5$ has the largest fitness value, which means it is the best example among all. The probability of $AS^t_5$ is selected as a parent should also be the largest, i.e. $0.21$. The probability sum of all examples is $1$.

\begin{table}[H]
\centering
\caption{Perturbations evaluation metrics.}
\label{my-label:Calculation results}
\begin{tabular}{l l l l l l l}
\hline
\hline
     & $AS^{t}_{1}$ & $AS^{t}_{2}$   & $AS^{t}_{3}$ & $AS^{t}_{4}$ & $AS^{t}_{5}$   & $AS^{t}_{6}$\\ \hline
$\phi(AS^{t}_{i})$  & 0.9& 0.45 & 0.6& 0.79&0.95&0.85 \\
$f(AS^{t}_{i})$  & 0.20& 0.10 & 0.13& 0.17&0.21&0.19 \\
$fr(AS^{t}_{i})$  & 0.20&0.30&0.43&0.60&0.81&1 \\ \hline
\hline
\end{tabular}
\end{table}

\begin{figure}[H]{
\centering
\captionsetup{justification=centering}
  \subfloat[The schematic of the crossover and variation.]{\includegraphics[width=0.5\linewidth]{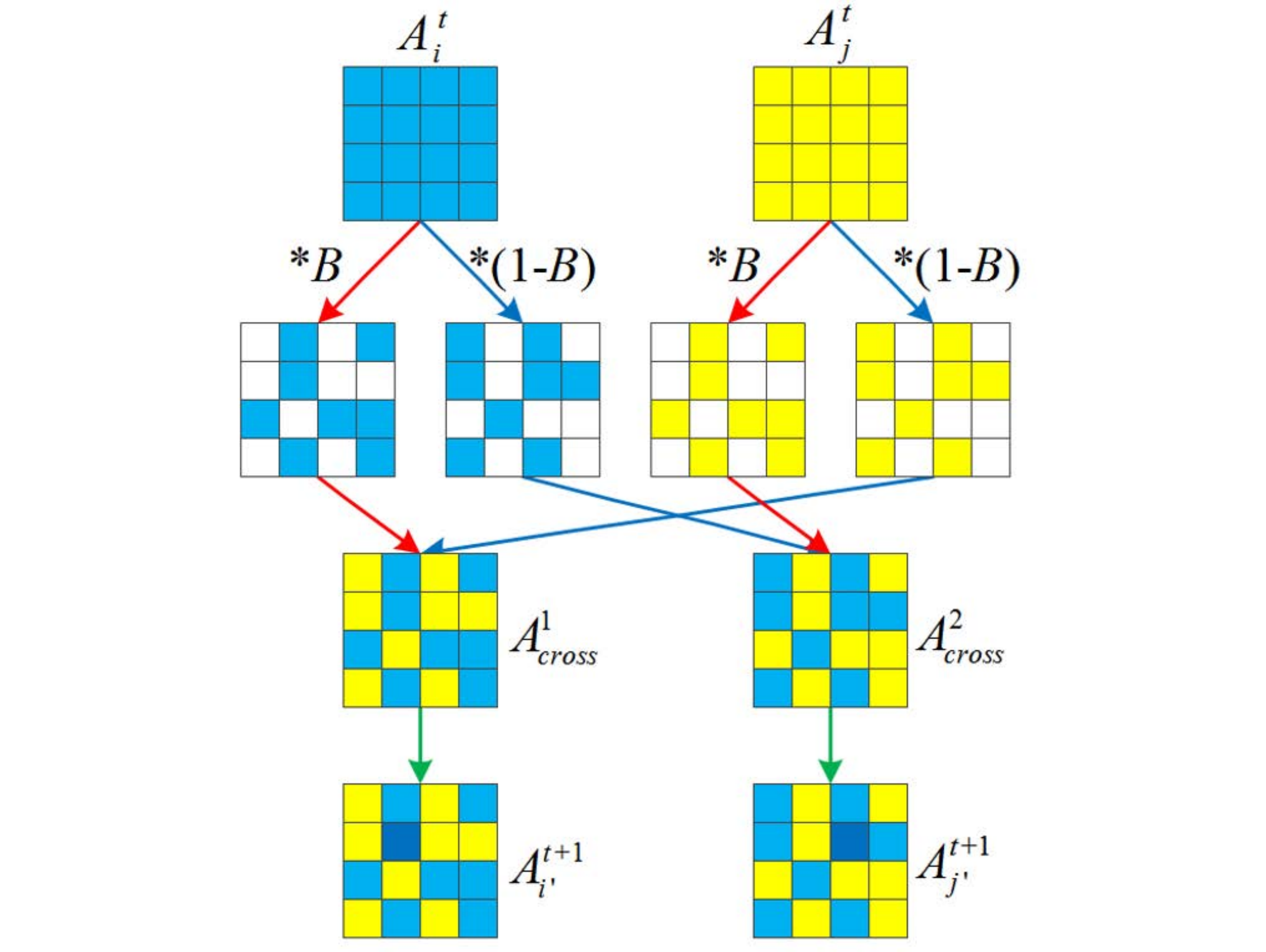}}\vspace{1ex}
  \subfloat[The specific example of crossover and variation.]{\includegraphics[width=0.5\linewidth]{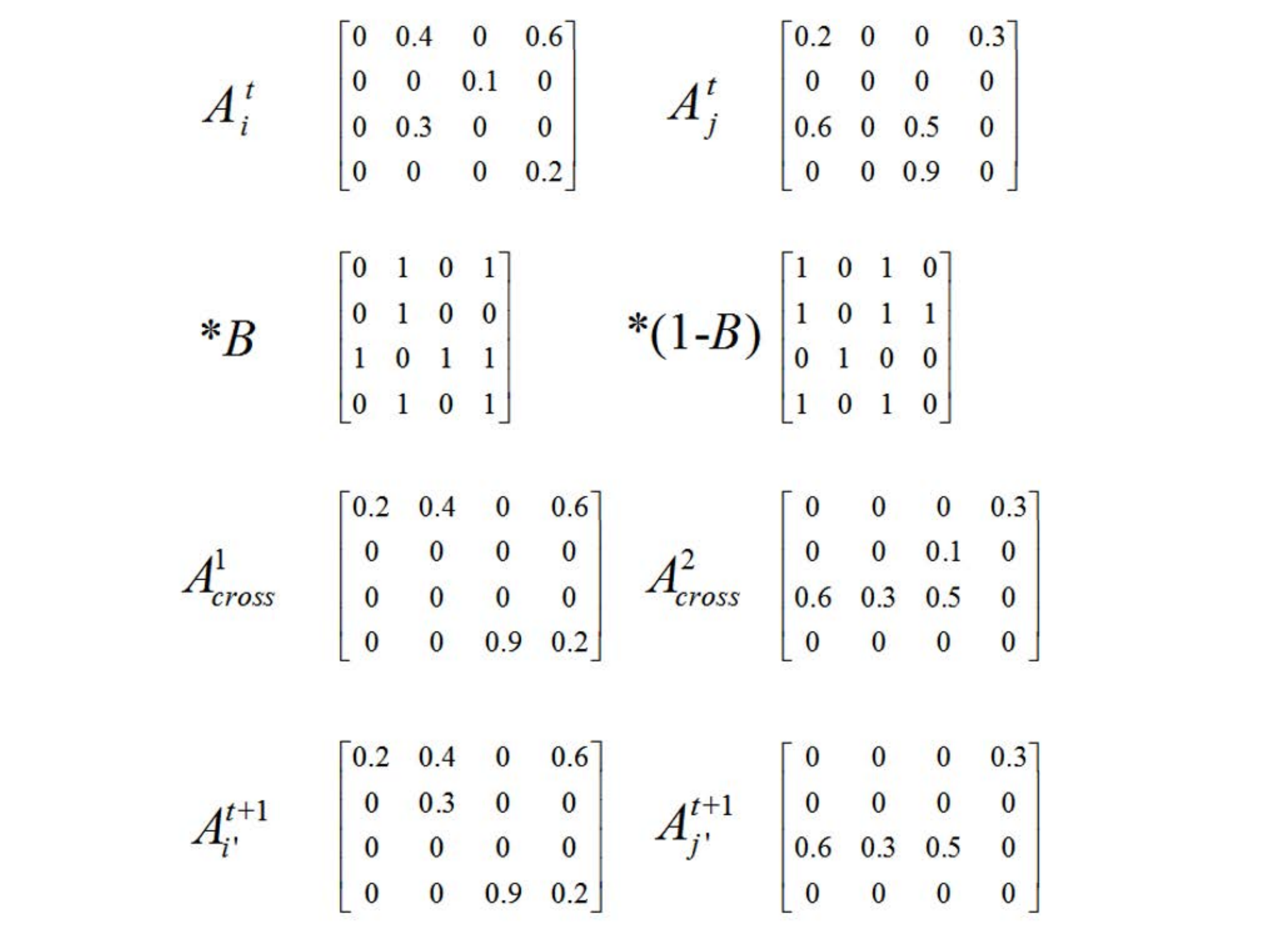}}\vspace{1ex}}

\caption{Crossing and mutating two parents to obtain new perturbations. The perturbation is divided into two complementary parts by the matrices $B$ and $1-B$, and then recombined with the complementary parts of the other perturbation to obtain two new perturbations.}
\label{fig:Crossover1}
\end{figure}

\subsubsection{Crossover operator}

Crossover operator is executed to generate new examples from selected examples. This paper takes a uniform crossover, in which the genes at each locus of two paired individuals are exchanged with the same crossover probability, thus forming two new individuals. Figure ~\ref{fig:Crossover1}(a) shows the schematic of the crossover and variation of the example. To make the crossover process clearer, we split the step into a 2-step display. Unlike traditional chromosome crossover, our perturbation crossover can be considered as two-dimensional matrices crossover. $A^t_i$ and $A^t_j$ are the two parent perturbations selected, represented in Figure~\ref{fig:Crossover1}(a) by blue and yellow matrices, respectively. Then, they are divided into two parts according to the same cross matrix. Where $B$ is a two-dimensional matrix, and the RGB pixel value of the position is exchanged when crossing. Finally, cross perturbation to generate new perturbations $A^{1}_{cross}$ and $A^{2}_{cross}$ to pass to the next step. This process can be represented by equations Eq.~\ref{equ:Eq_Crossover1_1} and Eq.~\ref{equ:Eq_Crossover1_2}. You can find a specific example in the figure ~\ref{fig:Crossover1}(b).

\begin{equation}
\begin{array}{c}
A^{1}_{cross}=\left\{
            \begin{array}{ll}
            A^{t}_{i}*B+A^{t}_{j}*(1-B) & rand(0,1)<P_{c} \\
            A^{t}_i & otherwise\\
            \end{array}
\right.
\label{equ:Eq_Crossover1_1}
\end{array}
\end{equation}
\begin{equation}
\begin{array}{c}
A^{2}_{cross}=\left\{
            \begin{array}{ll}
            A^{t}_{i}*(1-B)+A^{t}_{j}*B & rand(0,1)<P_{c} \\
            A^{t}_j & otherwise\\
            \end{array}
\right.
\label{equ:Eq_Crossover1_2}
\end{array}
\end{equation}
where $B$ is a matrix of the same size of the image, each element of $B$ is random number of $0$ or $1$. $rand(0,1)$ means randomly generating a number between 0 and 1. $P_c$ represents the probabilities of crossover.
Figure~\ref{fig:p_c} in the appendix shows the effect of $P_c$ on experimental results. The experimental results show that the larger the $P_c$ is, the faster the fitness function converges. This is mainly because we use the parent-child hybrid method to generation updates and our mutation probability is very low. When $P_c<1$, it is very likely to generate the same child example as the father, that is, generate duplicate examples, which increases the number of unnecessary queries and increases attack time and cost. Therefore, in order to reduce the cost of attack, this paper makes $P_c = 1$.

\subsubsection{Mutation operator}
Mutation operator indicates that some examples will be altered by a certain probability during the breeding process. In POBA-GA, we adopt multi-point mutation. According to mutation probability $Pm$, randomly select several pixels of $AS^{q}_{cross}$, where $q=\{1,2\}$. The effect of variation is shown in Figure~\ref{fig:Fig1}, and the specific process is shown in Figure~\ref{fig:Crossover1}. The process is defined as Eq.~\ref{equ:Eq_Evolutionary}.

\begin{equation}
\begin{array}{c}
A^{t+1}_{i'+q-1}=\left\{
            \begin{array}{ll}
            A^{q}_{cross}*C & rand(0,1)<P_{m} \\
            A^{q}_{cross} & otherwise\\
            \end{array}
\right.
\label{equ:Eq_Evolutionary}
\end{array}
\end{equation}
where $C$ is a matrix of the same size of the image. Except for a small number of elements in C, which are between 0 and 2, the remaining elements are all 1. $P_m$ represents the probability of mutation. Through experiments we make $P_m=0.001$ for the ImageNet data set and $P_m=0.003$ for the MNIST and CIFAR-10 data set.

Algorithm~\ref{algorithm: algorithm1} shows the pseudo code of POBA-GA, and Algorithm~\ref{algorithm: algorithm2} shows the pseudo code of the function in POBA-GA.

    \begin{algorithm}
        \caption{POBA-GA.}
        \label{algorithm: algorithm1}
        \begin{algorithmic}[1]
            \Require Input original example $S$, target model $TM$, parameter $P_{c}$, $P_{m}$, $\alpha$, maximum number of iterations $T$, population size $N$.
            \Ensure Approximate optimal adversarial example {$AS_{opt}$}
            \For{$i =1 \to N$}
                \State $S^{t=0}_{i++} \gets$ random initialization
            \EndFor
            \For{$t = 1 \to T$}
                \State $AS^{t} \gets A^{t}+S$
                \For{$i = 1 \to N$}
                \If {$y_{1}\ne y_{0}$}
                  \State $\phi(AS^{t}_{i})\gets p(y_{1}|AS^{t}_{i})-p(y_{0}|AS^{t}_{i})-\frac{\alpha}{\max Z(A^{t_0})} Z(A^{t}_{i})$
                \Else
                    \State $\phi(AS^{t}_{i}) \gets p(y_{2}|AS^{t}_{i})-p(y_{0}|AS^{t}_{i})$
                \EndIf
                \EndFor
                \For{$i = 1 \to N$}
                    \State $f(AS^{t}_{i}) \gets \frac{\phi(AS^{t}_{i})}{\sum_{j=1}^{n}\phi(AS^{t}_{j})}$
                    \State  $fr(AS^{k}_{i++}) \gets {\sum_{j=1}^{i}f(AS^{t}_{j})}$
                \EndFor
                \If{ max\{$\phi(AS^{t}_{i})\} > \gamma$}
                    \State \textbf{Break}
                \EndIf

                \For{$n = 1 \to N/2$}
                    \State $A^{t}_{i} \gets$ \Call{Selection}{$AS^{t}$}
                    \State $A^{t}_{j} \gets$ \Call{Selection}{$AS^{t}$}
                    \State $A^{t+1}_{2n-1}, A^{t+1}_{2n} \gets$ \Call{Crossover}{$P_{c}, A^{t}_{i}, A^{t+1}_{j}$}
                    \State $A^{t+1}_{2n-1} \gets$ \Call{Mutation}{$P_{m}, A^{t+1}_{2n-1}$}
                    \State $A^{t+1}_{2n} \gets$ \Call{Mutation}{$P_{m}, A^{t+1}_{2n}$}
                \EndFor
            \EndFor
            \State ${AS_{opt}} \gets$ the $AS_{i}^{t}$ with higerst $\phi(AS^{t})$
        \end{algorithmic}
    \end{algorithm}

    \begin{algorithm}
        \caption{The function in POBA-GA.}
        \label{algorithm: algorithm2}
        \begin{algorithmic}[1]
            \Function{Selection}{$AS^{t}$}
                \State $k \gets range(0,1)$
                \State $i \gets 1$
                \While{$k < fr(AS^{t}_{i})$}
                    \State $i++$
                \EndWhile
            \State \Return{$A^{t}_{i}$}
            \EndFunction

            \State
            \Function{Crossover}{$P_{c}, A^{t}_{i}, A^{t+1}_{j}$}
                \If {$rand(0,1) < P_{c}$}
                    \State $A^{t+1}_{i} \gets A^{t}_{i}*B+A^{t}_{j}*(1-B)$
                    \State $A^{t+1}_{j} \gets A^{t}_{i}*(1-B)+A^{t}_{j}*B$
                \Else
                    \State $A^{t+1}_{i} \gets A^{t}_{i}$
                    \State $A^{t+1}_{j} \gets A^{t}_{j}$
                \EndIf
                \State \Return{$A^{t+1}_{i}, A^{t+1}_{j}$}
            \EndFunction

            \State
            \Function{Mutation}{$P_{m}, A^{t+1}_{i}$}
                \If {$rand(0,1) < P_{m}$}
                    \State $A^{t+1}_{i} \gets A^{t+1}_{i}*C$
                \Else
                    \State $A^{t+1}_{i} \gets A^{t+1}_{i}$
                \EndIf
                \State \Return{$A^{t+1}_{i}$}
            \EndFunction
        \end{algorithmic}
    \end{algorithm}

\subsection{Generation update}
Generation is updated by father-son mixed selection. The population size $N$ perturbation with the highest fitness function in $A^t_i$ and $A^{t+1}_i$ are updated to $A^{t+1}_i$. This method is mainly used to prevent the optimal individual of the current group from being lost in the next generation, which causes the genetic algorithm cannot converge to the global optimal solution. And the termination condition of this paper is to achieve the number of cycles or the fitness function is greater than a certain value $\gamma$.

\section{Evaluation Metrics\label{Evaluation Metrics}}
{
In general, the researchers use $L_0$, $L_2$ and $L_{\propto}$ to calculate the perturbation size. However, this paper finds that there are problems in these three metric, so this paper proposes a perturbation evaluation metric improved from Sigmoid. When evaluating perturbation by the naked eye, our metric is more in line with visual assessment of perturbation. When evaluating perturbation by machine, our metric can speed up the reduction of perturbation in adversarial attack, and effectively reduce the difference between adversarial examples and original examples.

\subsection{Naked-eye evaluation}
Current adversarial attack methods often use p-norm $L_{p}$ to limit perturbations[1] of adversarial examples. For example, the zero-norm $L_{0}$ distance measures the number of pixels that have changed, the second-norm $L_{2}$ distance measures the Euclidean distance between two images, and the infinite-norm $L_{\propto}$ distance measures the maximum perturbation between two pixels.

\begin{figure}[H]
\centering
\includegraphics[width=0.7\linewidth]{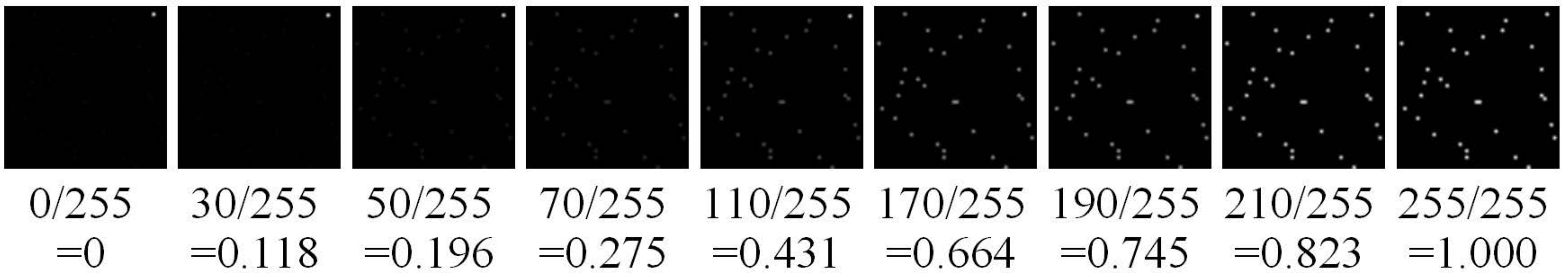}
\caption{Perturbations with different pixel values of noise points. Change the 30 pixels on the $50*50$ image as perturbation and normalize the changed pixel value.}
\label{fig:noise}
\end{figure}

However, no matter what kind of metric we use, we can not express the relationship between the magnitude of the nine perturbations in Figure~\ref{fig:noise}. $L_{0}$ and $L_{\propto}$ can't compare the perturbation size of these 9 pictures, and $L_{2}$ thinks that the difference between the $9^{th}$ picture and the $7^{th}$ picture is equal to the difference between the $1^{st}$ picture and the $5^{th}$ picture. However, it is difficult to distinguish the difference between the last three pictures or the difference between the first two pictures, and the difference between the $5^{th}$ and $6^{th}$ picture can clearly distinguished.

The calculation of the perturbation size is more similar to the sigmoid curve distribution. Inspired by this function, we propose a new perturbation calculation indicator $Z(A^{t}_{i})$. For a given adversarial example $A^t_{i}$, the perturbation is by Eq.~\ref{equ:Eq_7}.

\begin{equation}
\begin{array}{c}
Z(A^{t}_{i})=\sum_{a=1}^{m_a}\sum_{b=1}^{m_b}(\frac{1}{1+e^{-|AS^{t}_{i(ab)}|*pm_1+pm_2}}-\frac{1}{1+e^{pm_2}})
\label{equ:Eq_7}
\end{array}
\end{equation}
where $m_a$ and $m_b$ are the height and width of the image. $AS^{t}_{i(ab)}$ is the perturbation pixels in row $a$ and column $b$. $pm_1$ and $pm_2$ are the parameters to adjust the perturbation pixel mapping rule. When we evaluate with the naked eye, we set $pm_1=10$, $pm_2=5.8$.

Compare new perturbation evaluation metric with the $L_2$, Figure~\ref{fig:perturbation Z} is a comparison of perturbations evaluation metrics for a single perturbed pixel. Among them, the red line is the calculation result of the perturbations evaluation of our metric, and the blue line is the perturbations evaluation of the $L_{2}$. Combined Figure~\ref{fig:noise} and Figure~\ref{fig:perturbation Z}, we can find that $Z(A^{t}_{i})$ is more in line with the sensitivity of the human eye to perturbation. The metric proposed in this paper can effectively compensate for the vulnerabilities of $L_0$, $L_2$ and $L_{\propto}$ .

\begin{figure}[H]
\centering
\includegraphics[width=0.55\linewidth]{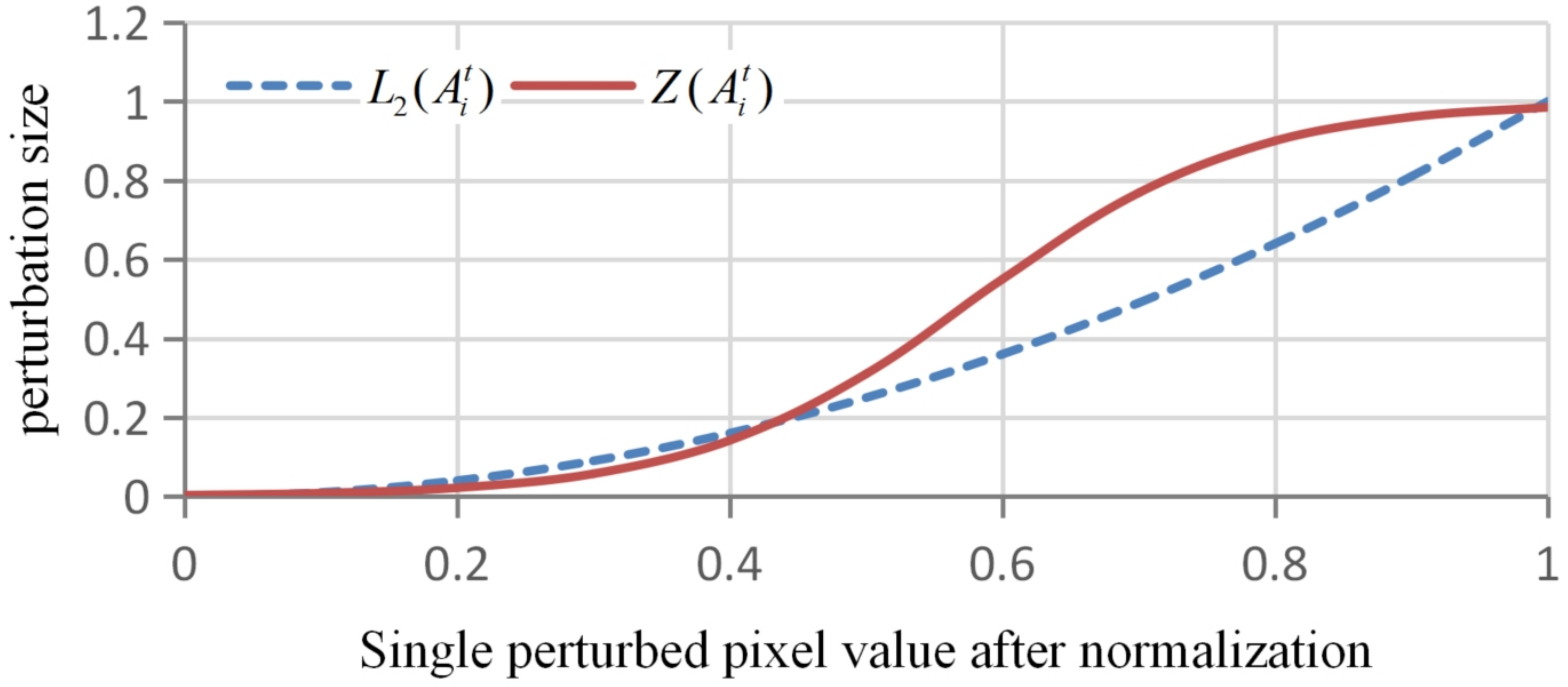}
\captionsetup{justification=centering}
\caption{Perturbation evaluation curves of $L_2$ and $Z$.}
\label{fig:perturbation Z}
\end{figure}

\subsection{Machine evaluation}
In the last section we have shown that $Z(A^{t}_{i})$ is more in line with the sensitivity of the human eye to perturbation. And in this section we mainly prove that $Z(A^{t}_{i})$ can better distinguish small perturbation, making the perturbation of adversarial examples generated by adversarial attacks smaller. In general, the maximum perturbation is less than $0.43(110/255)$, this paper set $pm_1=15$, $pm_2=3$.

In order to compare the impact of using the $Z(A^{t}_{i})$ metric and the $L_2$ metric on the POBA-GA attack. We set the fitness function to $\phi(AS^{t}_{i})= P(AS^{t}_{i})- \frac{\alpha}{\max Z(A^{0})} Z(A^{t}_{i})$
and  $\phi(AS^{t}_{i})= P(AS^{t}_{i})- \frac{\alpha}{\max ||A^{0}_{i}||_2} ||A^{t}_{i}||_2$ respectively. The experimental settings refer to section~\ref{Exp}, and iterating 100 generations. The experimental results are shown in Table~\ref{my-label:l2ad}. It should be noted that in order to prove the validity of our metric, we use $L_2$ to calculate the perturbation of adversarial examples. Table~\ref{my-label:l2ad} shows that our metric can speed up the reduction of perturbation in adversarial attack, and effectively reduce the difference between adversarial examples and original examples. This is mainly because $Z(A^{t}_{i})$ expands the difference between perturbation and makes the perturbation optimize faster in the direction of smaller perturbation. From the table we can find that their attack success rate is 96\%, because we do not consider the perturbation before the attack is successful, so the metric does not affect the attack success rate.

\begin{table*}[!ht]
\centering
\caption{Influence of disturbance metric on POBA-GA in VGG19}
\label{my-label:l2ad}
\begin{tabular}{ c c c c}
\hline
\hline
        &   Attack success rate     &   Query cout &   Perturbation(per-pixel $L_2$)\\
        \hline
$L_2$     &   96\%                    &   5000   &5.7e-06\\
$Z$     &   96\%                    &   5000  & 4.3e-06\\
\hline
\hline
\end{tabular}
\end{table*}

}

\section{Experiments\label{Exp}}
{In this section, we separately experiment on the parameter sensitivity, attack performance, universality of the attack algorithm, its impact on the robustness of the model and the practical application of the algorithm. In 5.1, we describe the platform, database, DNN models and attack implementation, making it easy for readers to reproduce the experiment. In 5.2, we analyzed the effect of parameter $\alpha$ on the experiment, so that people can quickly select different parameters $\alpha$ according to their needs. In 5.3, we compared the POBA-GA with high-quality white-box and black-box attacks, including the state of the art white-box attack C\&W and black-box attack AutoZOOM~\cite{Tu2018AutoZOOM}. In 5.4, we study the experiment when the black-box model only return just a single binary outcome and target attack. And we want to show that our POBA-GA can achieve good attack effect and universality in such situation. In 5.5, we compare the defense performance of POBA-GA adversarial training and ensemble adversarial training, hoping to prove that the defense capability of the POBA-GA adversarial training is stronger than the ensemble adversarial training. In 5.6, we attack the face recognition system, hoping to prove that OPBA-GA not only has good experimental results in the experimental data set, but also can get good attack results in the real world and other data sets.

\subsection{Experiment setup}
\textbf{Platform:} The platform for all experiments is i7-7700K 4.20GHzx8 (CPU), TITAN Xp 12GiBx2 (GPU), 16GBx4 memory (DDR4), Ubuntu 16.04 (OS), Python 3.5, Tensorflow-gpu-1.3, Tflearn-0.3.2\footnote{Tflearn can be download at \emph{https://github.com/tflearn/tflearn}}.

\textbf{Database:} We use three publicly available image databases, namely, MNIST\footnote{MNIST can be download at \emph{http://yann.lecun.com/exdb/mnist/}}, CIFAR-10\footnote{CIFAR-10 can be download at \emph{ https://www.cs.toronto.edu/~kriz/cifar.html}}, ImageNet64\footnote{ImageNet64 can be download at \emph{http://image-net.org/download-images}}. MNIST is a handwritten digitally recognized data set containing 70,000 grayscale images of 28*28 size, divided into 10 classes. The CIFAR-10 data set is also a small image data set, contains 60,000 color images of size 32*32, divided into 10 classes. The Imagenet data set is a large image data set, contains about 15 million images and 22,000 classes. Before the experiment, we reshaped it to a size of 224*224.

 \textbf{Baseline methods:} We compare POBA-GA with eight different baselines, including four white box attacks and four black box attacks,including Fast Gradient Sign Model (FGSM)~\cite{goodfellow2014explaining}, DeepFool~\cite{moosavi2016deepfool}, Basic Iterative Model (BIM)~\cite{Kurakin2016Adversarial}, Carlini and Wagner Attacks (C\&W)~\cite{carlini2017towards}, Boundary~\cite{brendel2017decision}, ZOO~\cite{chen2017zoo}, AutoZOOM~\cite{Tu2018AutoZOOM}. These baselines are classic and efficient attack methods, including the state of the art white-box attack C\&W and black-box attack AutoZOOM.

\textbf{DNN Models:} For MNIST and CIFAR-10, we use the same DNN model as in the C\&W attack~\cite{carlini2017towards}\footnote{\emph{https://github.com/carlini/nn\_robust\_attacks}(commit 1193c79)}, which is also used by ZOO and Boundary~\cite{brendel2017decision}. On ImageNet we use the same pretrained networks VGG19~\cite{simonyan2014very}, Resnet50~\cite{he2016deep} and Inception-V3 (Inc-V3)~\cite{szegedy2016rethinking} provided by Keras \footnote{\emph{https://github.com/fchollet/keras}(commit 1b5d54)} as Boundary.

 \textbf{Attack implementation:} In order to make the experimental results of the comparative experiment accurate, the parameters of the comparison algorithm are taken directly from the corresponding literature. In this paper, POBA-GA performs 100 iterations on MINIST and CIFAR-10, generating 20 descendants per iteration, the variance is taken from $[5,10,15,20,25]$, the number of noise points is taken from $[50,100,150,200,250]$. And POBA-GA performs 400 iterations on ImageNet, and generating 50 descendants per iteration, the variance is taken from $[5,10,15,20,25]$, the number of noise points is taken from $[5000,7500,10000,12500,15000]$.

 All experimental results in this paper are average values. For MNIST and CIFAR-10, we evaluated 1000 randomly examples from the validation set, for ImageNet we used 250 images.

\textbf{Evaluation metric:} This paper uses the attack success rate (ASR)~\cite{Dong2017dongboosting}, perturbation (per-pixel $L_2$) and query number to evaluate the performance of the experiment. The ASR is used to evaluate the attack probability of the attack method against the target model, which is calculated by Eq.~\ref{equ:Eq_attack success rate}.

\begin{equation}
\begin{array}{c}
ASR=\left\{
            \begin{array}{ll}
            \frac{sumNum({AS_{opt}}|y_{1}=y_{tar})}{sumNum({AS_{opt}})} & $targeted attack$\\
            \frac{sumNum({AS_{opt}}|y_{1}\neq{y_{0}})}{sumNum({AS_{opt}})} & $non-targeted attack$
            \end{array}
\right.
\label{equ:Eq_attack success rate}
\end{array}
\end{equation}
where $sumNum(.)$ is the number of examples.

$L_2$ norm (i.e. Euclidean distance) is used to quantify the difference between the adversarial and the original examples. Query number refers to how many examples the target model needs to predict before the attack method reaches the stop condition. This metric is especially important when the target model has a limit on the number of queries.

}

\subsection{Influence of perturbation ratio parameter $\alpha$}

The perturbation weight $\alpha$ is used to adjust the balance between perturbation $Z(A^t_i)$ and attack performance $P(AS^t_i)$. When $\alpha$ is larger, the example is more similar to the original one, but the success rate will be lower. When $\alpha$ is smaller, the perturbation of the adversarial example is larger, the attack success rate is higher. {The settings of parameter $\alpha$ will change as people's attack requirements change, so we analyzed the effect of parameter $\alpha$ on the experiment so that people can quickly select different parameters $\alpha$ according to their needs. In addition, we also pointed out in the article that if people have clear expectations, they can also use automatic methods to adjust the parameter $\alpha$, such as irace\footnote{Irace \emph{https://cran.r-project.org/web/packages/irace/index.html}}}

\begin{figure}[H]{
\captionsetup{justification=centering}
  \subfloat[Perturbation $Z$.]{\includegraphics[width=0.48\linewidth]{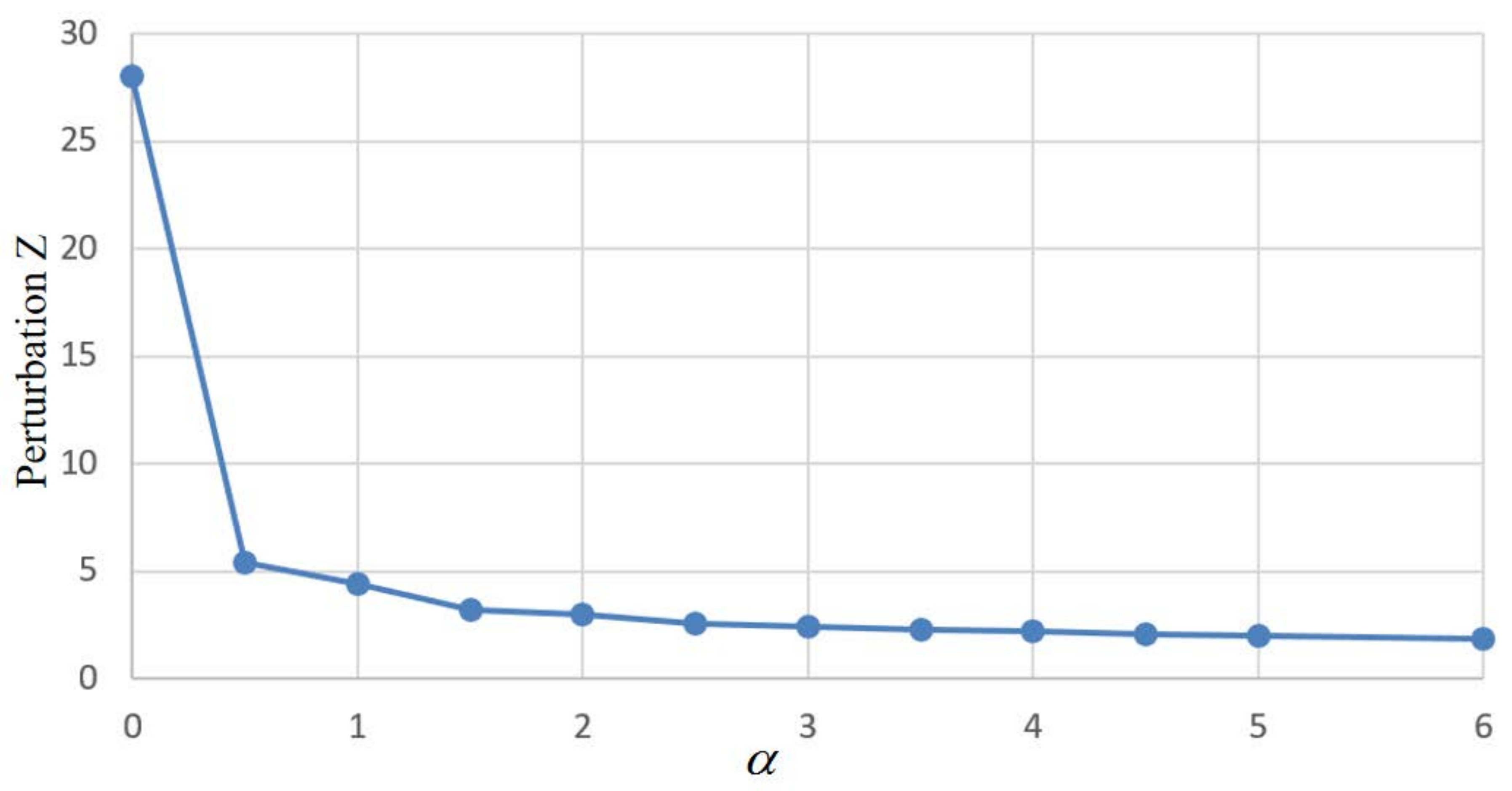}}\vspace{1ex}
  \subfloat[Attack Performance $P$]{\includegraphics[width=0.51\linewidth]{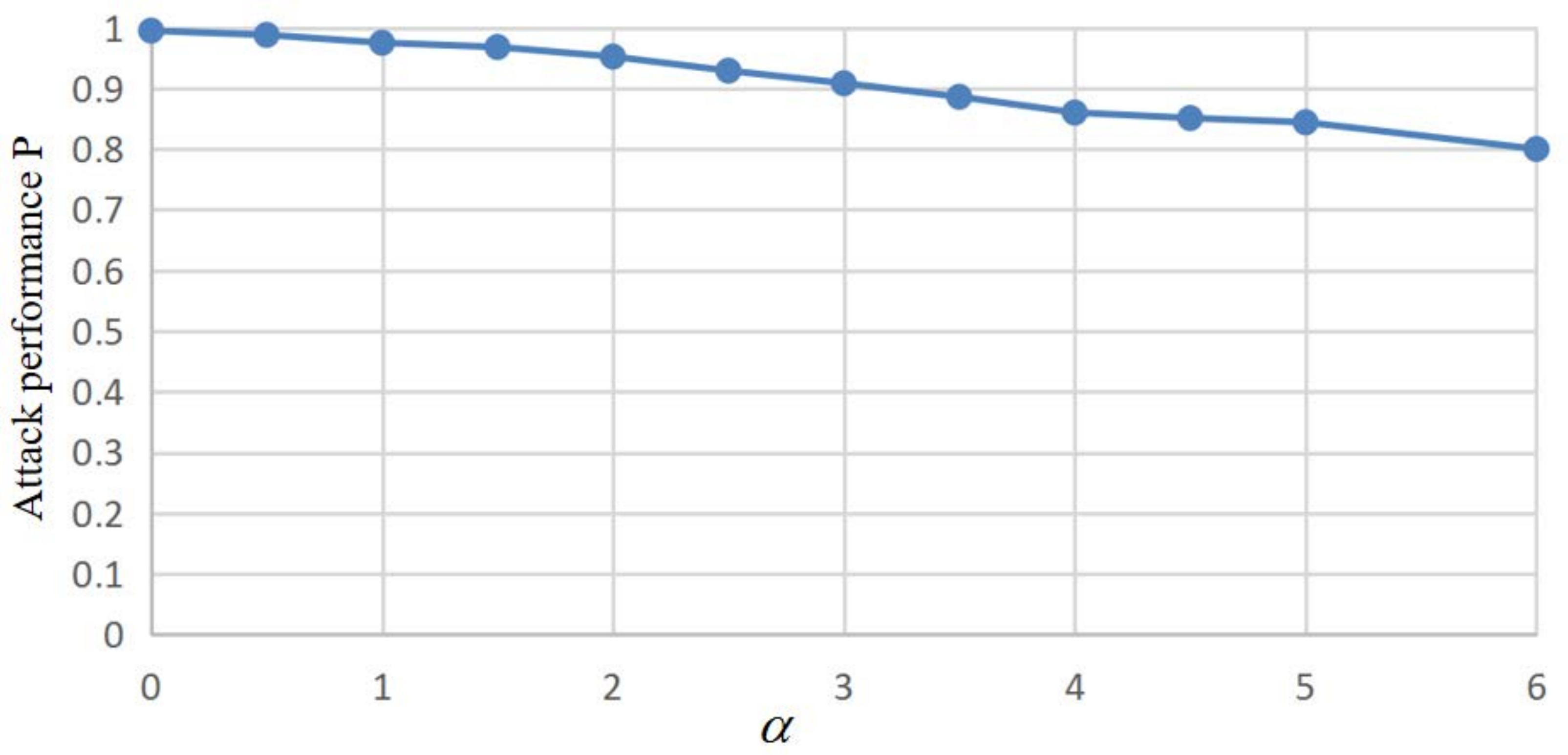}}\vspace{1ex}}
  \captionsetup{justification=centering}
\caption{The influence of $\alpha$ on the ImageNet64 data set.}
\label{fig:Fig5}
\end{figure}

Figure~\ref{fig:Fig5} shows the influence of $\alpha$ on the ImageNet64 data set. From the figures, we can conclude that with the increase of $\alpha$, the attack performance $P$ and perturbations $Z$ are reduced. This is mainly because with the increase of $\alpha$, the influence of perturbation on the image gradually increases, the adversarial example is more similar to the original image, and the attack performance $P$ gradually decreases.

According to different actual needs, we will choose a different $\alpha$. In this paper, we makes $\alpha = 3$, because the perturbation almost does not decrease with the increase of $\alpha$, and $P$ is still as high as 0.9. It should be noted that although $\alpha$ affects the value of $P$, since we consider the perturbation after the attack is successful, in general, $\alpha$ does not affect the attack success rate.

\subsection{Attack performance comparison}
{To verify the effectiveness of our approach, we compared POBA-GA with the classic white-box and black-box attack methods. We found that POBA-GA has a high attack success rate, even surpassing some white-box attacks, and it can initially achieve successful attacks with a small number of queries.} Table~\ref{my-label:perturbation} shows the attack success rate, perturbations and attack time cost of the different adversarial attacks.{ The number of queries in parentheses represents the number of queries when the initial attack succeeds, which is used to assist in comparing the performance of our algorithm with ZOO and AutoZOOM algorithms.}

\subsubsection{Perturbation} The adversarial examples generate by POBA-GA has less perturbation with the same attack success rate. Especially for MNIST and CIFAR-10 datasets, examples generated by our method have less perturbation than most white-box and black-box attacks,{ mainly because the image is small and the genetic algorithm is easier to find the approximate global optimal solution. For the ImageNet64 data set, since the white-box attack method grasps the internal structure of the target model, we only compare with the black-box attack. From Table~\ref{my-label:perturbation} we find that the perturbation of POBA-GA is significantly smaller than ZOO and AutoZOOM. Although POBA-GA's perturbation is larger than Boundary, POBA-GA has fewer queries. In addition, when $\alpha=10$ and the number of queries is 54000, the perturbation generated by POBA-GA can reach 8.1e-07 without affecting the attack success rate, but the attack perturbation $P$ will decrease. Therefore POBA-GA performs well in perturbation performance. It makes good sense that the perturbations of POBA-GA are much smaller than baselines since the initial perturbations are very small, and the optimization process of GA is capable of gradually reducing the perturbation during iterations.}

\subsubsection{Attack success rate}{POBA-GA has a better attack success rate and even better than some white box attacks. Usually, white-box attack conducted based on the internal structure and parameters of the model, can perform a higher success rate attack. The attack success rate of DeepFool is low, mainly because it has strict requirements on perturbation. From Table ~\ref{my-label:perturbation} you can find that ZOO also has a higher attack success rate, however it mainly relies on a large number of queries and iterations.

Specifically, on the ImageNet64 data set, ZOO requires about 220 thousand queries for 90\% attack success rate, AutoZOOM requires about 1600 queries for 100\% attack success rate, while POBA-GA only need about 500 queries for 96\% attack success rat. In addition, the perturbation of POBA-GA is only one-fifth of the ZOO and AutoZOOM, or even smaller. On the whole, POBA-GA achieves state-of-the-art black-box attack performances in consideration of both attack success rate and perturbation. Although there is small margin in success rate when compared with AutoZOOM, POBA-GA needs much less query times and less perturbations. We can conclude that POBA-GA is more practical in real world application when the require time is strict while we can tolerate with relatively high (less than 100\%) attack success rate.

The reason why we have such a high attack success rate is mainly because we do not consider the effects of the perturbation on the experiment before the attack is successful. But it should be noted that even if we do not consider the effects of the perturbation, the perturbation will not be very large, as we have limited the perturbation at initialization.

}

\subsubsection{Query count} {POBA-GA has fewer queries than other black-box attacks. Since white-box attacks are conducted based on the internal structure of the model, we only analyze black-box attacks. From Table ~\ref{my-label:perturbation}, we can find that POBA-GA and AutoZOOM need significantly fewer queries than Boundary and ZOO. AutoZOOM reduces the query time by adopting an adaptive random full gradient estimation strategy to strike a balance between query counts and estimation errors, and features a decoder (AE or BiLIN) for attack dimension reduction and algorithm acceleration. POBA-GA reduces the number of queries by genetic algorithm. The main reasons are as follows: 1) When selecting examples, it is more likely to choose examples with high adaptability.  2)The high probability of crossover and variation increases the example diversity. 3) Use father and son mixed selection to retain the best example. 4)The effects of the perturbations are not considered before the attack is successful.
}

\begin{table*}[!t]
\centering
\caption{Attack performance comparison.}
\label{my-label:perturbation}
\begin{tabular}{ l c l c c c c c}
\hline
\hline
\multirow{2}{*}{}                                                                        & \multirow{2}{*}{Black/White-box} &\multirow{2}{*}{Attack method} & \multirow{2}{*}{MNIST} & \multirow{2}{*}{CIFAR-10} & \multicolumn{3}{c}{ImageNet64}                    \\ \cline{6-8}
                                                                                         &                            &                        &                        &   & VGG19          & Resnet50      & Inc-V3   \\ \hline
\multirow{8}{*}{\begin{tabular}[c]{@{}l@{}}Perturbation\\(per-pixel $L_2$)\end{tabular}}
& \multirow{4}{*}{\begin{tabular}[c]{@{}l@{}}White-box \\ Adversarial attack\end{tabular}} & FGSM                       & 6.5e-02                & 7.3e-05               & 3.4e-06        & 4.0e-06        & 2.6e-06        \\
                                                                                       &  & DeepFool                   & 3.2e-03               & 4.1e-06                &2.4e-07  & 9.3e-08 & 9.1e-08          \\
                                                                                        & & BIM                        & 8.2e-03                & 1.2e-05                 & 8.5e-07      & 8.2e-07         & 6.4e-07         \\
                                                                                         & & C\&W                       & 3.1e-03              & 6.9e-06            & 5.7e-07          & 2.2e-07      & 7.6e-08  \\ \cline{2-8}
& \multirow{4}{*}{\begin{tabular}[c]{@{}l@{}}Black-box \\ Adversarial attack\end{tabular}} & Boundary                   & 4.0e-03               & 6.4e-06            & 3.5e-07          & 2.1e-07         & 4.2e-07       \\
                                                                                     &    & ZOO                        & 4.3e-03               & 5.8e-04            & 3.9e-05          & 3.2e-05         & 2.8e-05     \\
                                                                                     &    & AutoZOOM                     & 6.4e-03               & 7.2e-04            & 6.2e-05          & 5.1e-05         & 5.4e-05     \\
                                                                                     &    & POBA-GA                      & 3.0e-03         & 6.8e-05         & 1.5e-05         & 1.4e-05       & 1.7e-05      \\ \hline
\multirow{8}{*}{\begin{tabular}[c]{@{}l@{}}Attack \\ success rate\end{tabular}}
&\multirow{4}{*}{\begin{tabular}[c]{@{}l@{}}White-box \\ Adversarial attack\end{tabular}}  & FGSM                       & 86\%                   & 89\%                   & 77\%          & 80\%           & 82\%          \\
                                                                                        & & DeepFool                   & 90\%                   & 87\%                   & 75\%          & 79\%           & 75\%          \\
                                                                                      &   & BIM                        & 98\%                   & 98\%                   & 97\%          & 96\%           & 94\%          \\
                                                                                        & & C\&W                       & 100\%         & 100\%         & 99\% & 100\% & 99\% \\ \cline{2-8}
&\multirow{4}{*}{\begin{tabular}[c]{@{}l@{}}Black-box \\ Adversarial attack\end{tabular}} & Boundary                   & 78\%                   & 76\%                   & 72\%          & 70\%           & 70\%          \\
                                                                                      &   & ZOO                        & 100\%        & 100\%         & 90\%          & 90\%           & 88\%          \\
                                                                                      &   & AutoZOOM                       & 100\%         & 100\%         & 100\%          & 100\%           & 100\%          \\
                                                                                      &   & POBA-GA                     & 100\%         & 100\%         & 96\%       & 98\%  & 95\% \\ \hline

\multirow{4}{*}{\begin{tabular}[l]{@{}l@{}}Query count\end{tabular}}

&\multirow{4}{*}{\begin{tabular}[c]{@{}l@{}}Black-box \\ Adversarial attack\end{tabular}} & Boundary                   & 12500                   & 12500                & 125000        & 125000          & 125000      \\
                                                                                       &  & ZOO                        & 9250                & 4324                 & 235272        & 223143        & 264170        \\
                                                                                       &  & AutoZOOM                        & 445(100)                 & 103(86)                 & 4647(1686)        & 4256(1695)        & 4051(1701)        \\
                                                                                      &   & POBA-GA     & 423(94)                & 381(78)                 & 3786(536)        & 3614(492)        & 3573(471)        \\ \hline
                                                                                      \hline
\end{tabular}
\end{table*}

\subsection{Universality}
\begin{figure}[H]
\centering
  \includegraphics[width=0.35\linewidth]{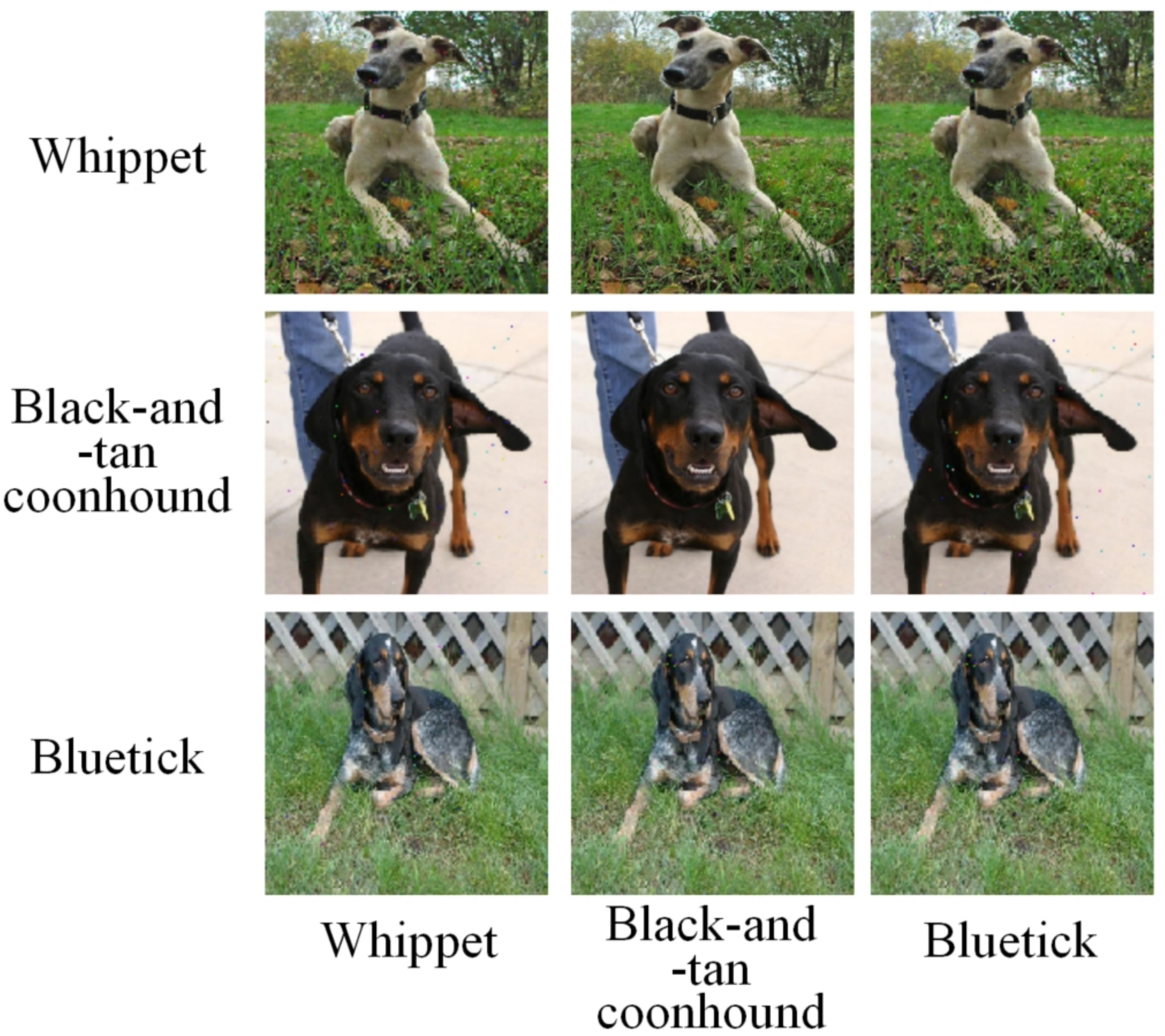}
\caption{Target attack. The rows represent the original label and the column represents the output label of the target model.}
\label{fig:Fig9}
\end{figure}

 \textbf{Targeted attack:}
Non-targeted attack only needs to make the $TM$ output label different from the correct one, and targeted attack need the target model output specified label. In order to verify the {universality} of the POBA-GA algorithm, we also achieve a target attack on $TM$. The fitness function value is designed as  $\phi(AS^{t}_{i})=p(y_{tar}|AS^{t}_{i})-p(y_{0}|AS^{t}_{i})-\alpha Z(AS^{t}_{i})$, where $p(y_{tar}|AS^{t}_{i}$) is the probability that the target model will classify the input picture as label $y_{tar}$. Figure~\ref{fig:Fig9} shows the targeted attack. Each line represents the original label, and each column represents the output label of the target model. The adversarial example average fitness function $\phi(AS_{pot})=0.53$, perturbation $Z(A_{pot})=3.846$. Therefore, the POBA-GA method can implement targeted attacks. We randomly selected 50 examples from the ImageNet64 data set to attack VGG19, inc-V3 and Resnet50. The attack success rates were $82\%$, $80\%$ and $84\%$, respectively.

{
 \textbf{A single binary outcome:} In general, the DNN model will give the confidence of the classification, but there are also some depth models that only return a single binary. Therefore, we experiment with models that just a single binary outcome. Different with RCC(Return confidence of classification), we cannot calculate the attack performance $P(AS^t_i)$ of RSB (Return a single binary). Therefor, we use Monte Carlo approximation to estimate the confidence of the classification, and then optimize the perturbation with POBA-GA. Estimating the confidence of $AS^t_i$ with Monte Carlo approximation can be expressed by the Eq.~\ref{equ:Eq_md}.

\begin{equation}
\begin{array}{c}
\widehat{P}(AS^{t}_{i})=\frac{1}{N'}\sum_{i=1}^{N'}R(AS^{t}_{i}+\delta)
\label{equ:Eq_md}
\end{array}
\end{equation}
where $\delta\sim\aleph(0,30)$, $N'=100$ is the number of examples used to estimating the confidence of $AS^{t}_{i}$. $R(AS^{t}_{i}+\delta)$ is the binary outcome of $AS^{t}_{i}+\delta$, for example, when $AS^{t}_{i}+\delta$ is predicted to the be the second class, then $R(AS^{t}_{i}+\delta)=[0,1,0,...,0]$.

\begin{table*}[!ht]
\centering
\caption{The performance of BOPA-GA on different outcome models}
\label{my-label:outcome}
\begin{tabular}{ c c c}
\hline
\hline
        &   Attack success rate     &   Query cout(initial success)\\
        \hline
RCC     &   98\%                    &   536\\
RSB     &   73\%                    &   6276\\
\hline
\hline
\end{tabular}
\end{table*}

Table ~\ref{my-label:outcome} shows the comparison using the RCC(Return confidence of classification) and RSB (Return a single binary) on the VGG19. From the table~\ref{my-label:outcome} we can find that if the model just return a single binary, the number of queries will increase by about 100 times, mainly because we need to estimate the confidence of the example through 100 queries. The attack success rate will be reduced from 98\% to 73\%. This is mainly because, although we have estimated the confidence by Monte Carlo approximation, there is a certain difference from the actual situation. If you want to increase the attack success rate, you can increase the $N'$ or the number of iterations.

\textbf{Real world experimentation:} In this section, we perform real world experimentation on GCP(which has a black box model), to demonstrate the applicability of the proposed methodologies. Considering the limited number of model queries in the actual scene, this paper only queries 1000 times, i.e. iterative 50 generations, generating 20 examples per generation. The experimental results are shown in Figure~\ref{fig:GCP}. Figure~\ref{fig:GCP}(a) is the prediction result of the original example, Figure~\ref{fig:GCP}(b) is the prediction result of a certain initial adversarial example, and Figure~\ref{fig:GCP}(c) is the prediction result of a $50^{th}$ generation adversarial example. The experimental results show that if we do not consider the perturbation, we can quickly achieve the attack on GCP. During the optimization process, we optimized the perturbation while ensuring that the attack was successful, and increased the confidence of the eggs from $64\%$ to $83\%$. Considering the limited number of model queries in the actual scene, this paper only queries 1000 times. If the query continues, the confidence of the real class label will further decrease. Therefore, POBA-GA can achieve the attack on the real world experimentation of GCP and proves the applicability of the proposed method.

\captionsetup[figure]{singlelinecheck=off}
\begin{figure}[H]
\centering
\captionsetup{justification=centering}
  \subfloat[The prediction result of the original example]{\includegraphics[width=0.48\linewidth]{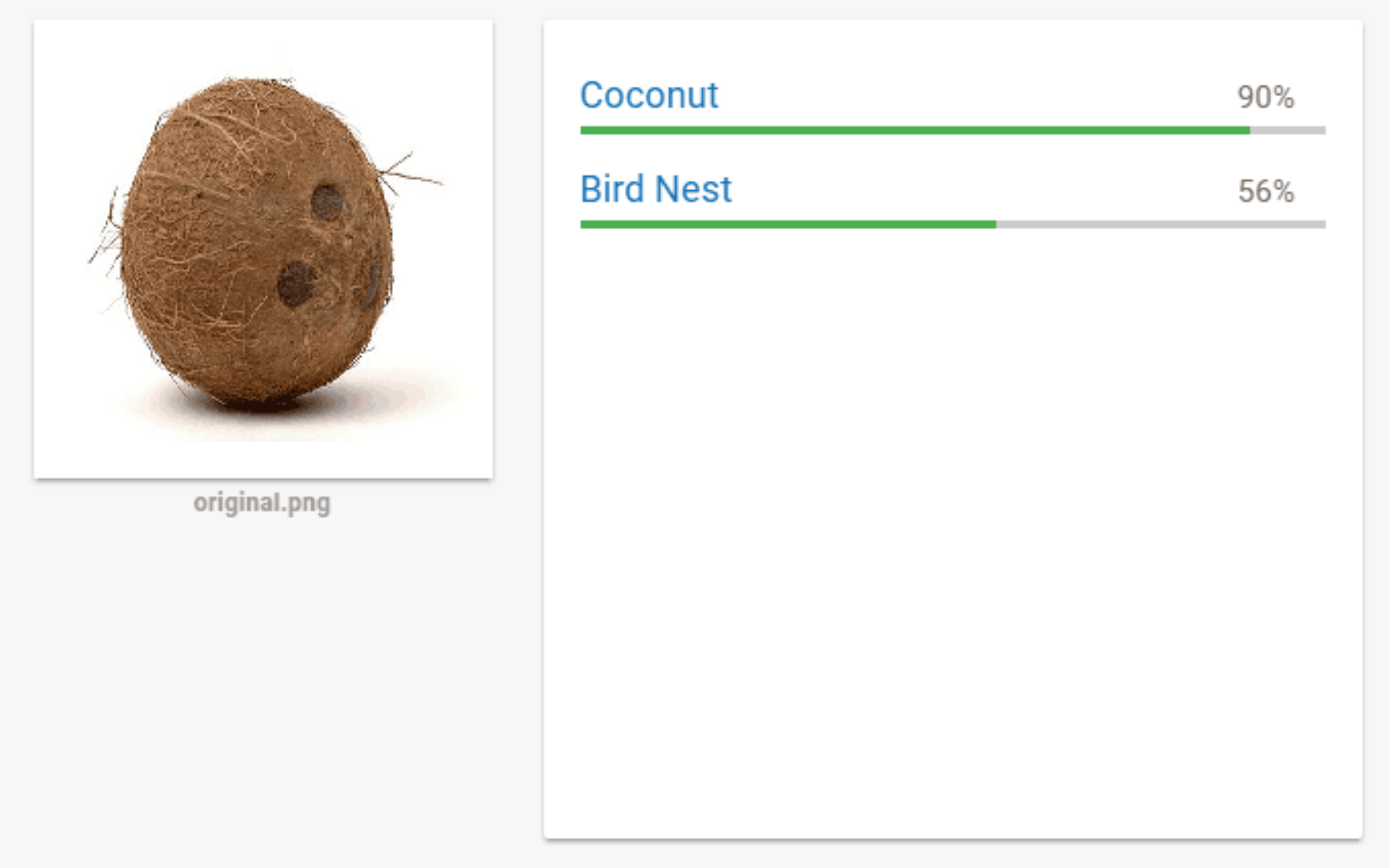}}\vspace{1ex}
  \subfloat[The prediction result of an initial adversarial example]{\includegraphics[width=0.48\linewidth]{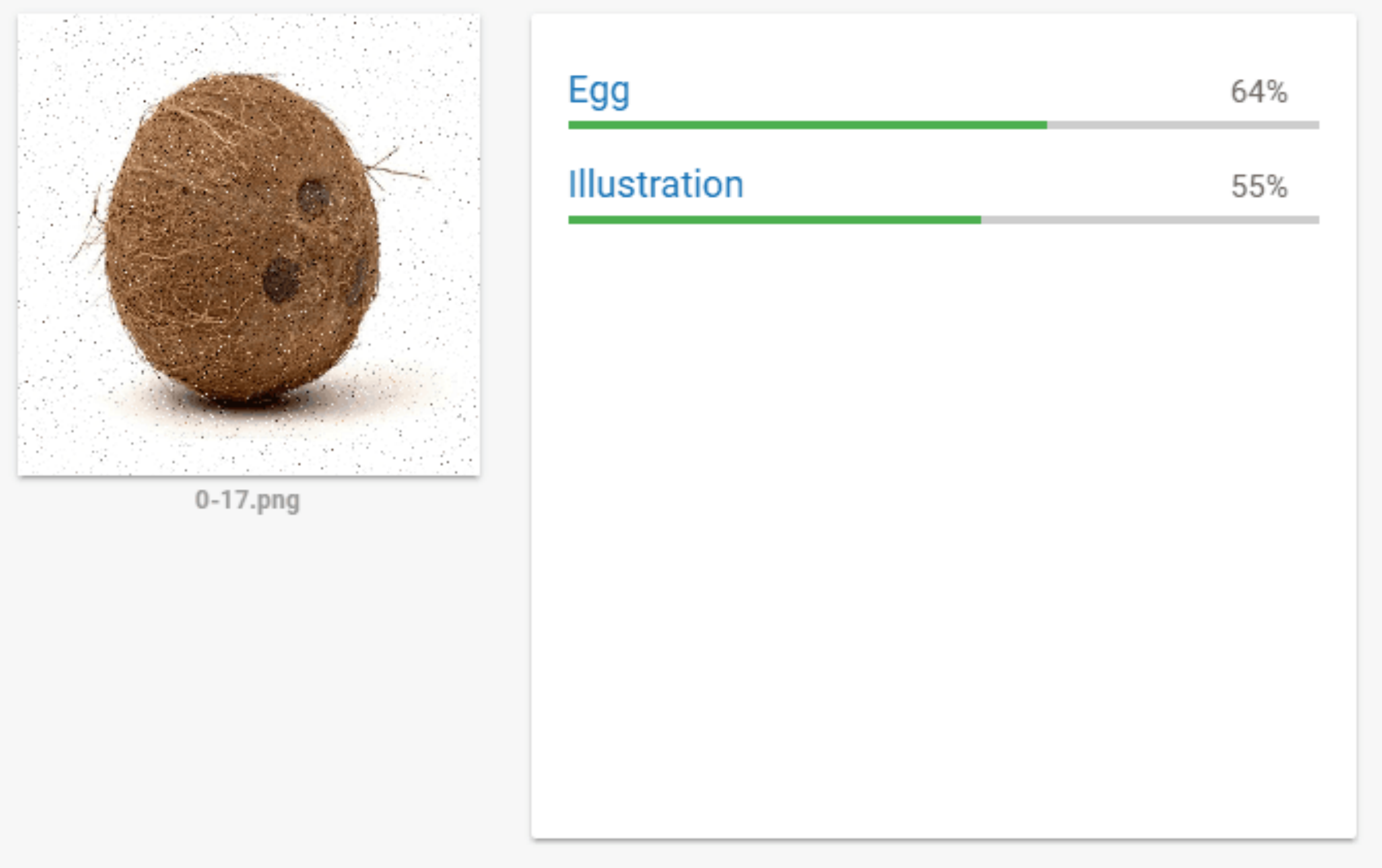}}\vspace{1ex}
  \subfloat[The prediction result of a $50^{th}$ generation adversarial example]{\includegraphics[width=0.48\linewidth]{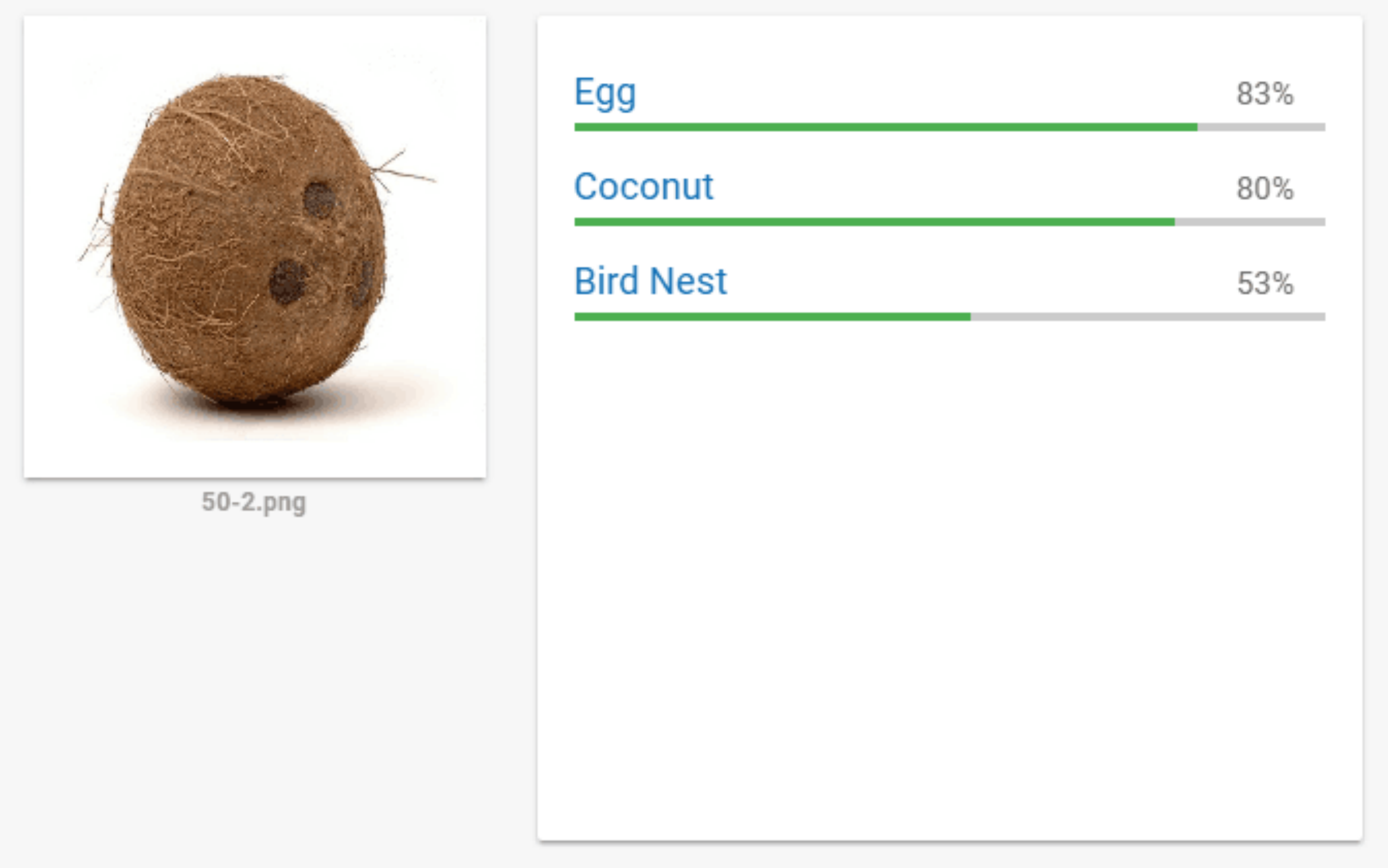}}\vspace{1ex}
\caption{The prediction results of GCP platform.}
\label{fig:GCP}
\end{figure}

}

\subsection{Defensiveness from POBA-GA adversarial training}
\begin{figure}[H]
\centering
\includegraphics[width=0.4\linewidth]{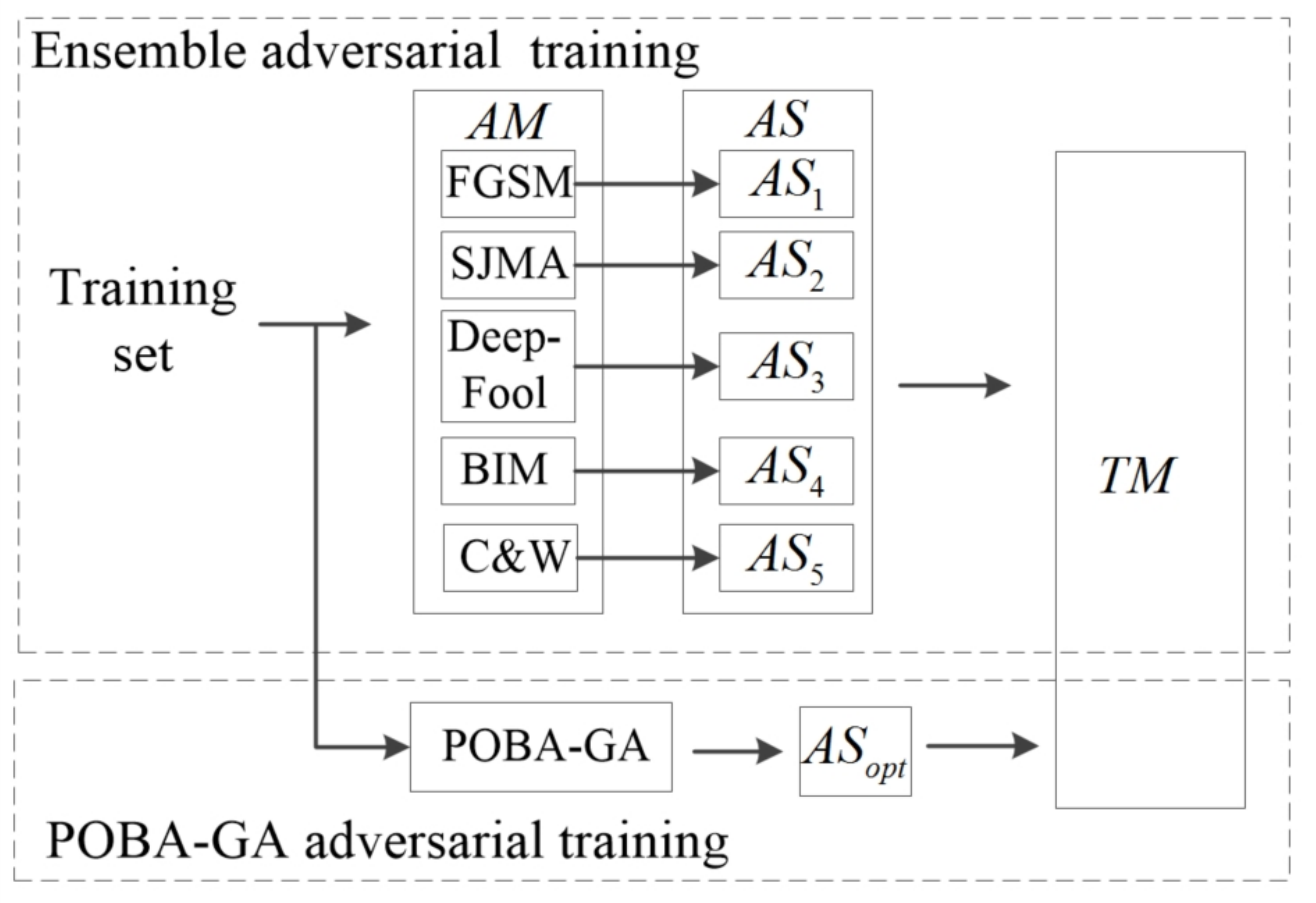}
\caption{The block diagram of the defense. Ensemble adversarial training is mainly used to train and defend against the adversarial examples generated by FGSM, SJMA, DeepFool, BIM and C\&W. POBA-GA adversarial training is the defense method proposed in this paper.}
\label{fig:defence}
\end{figure}

{
In this section, we compare the defense performance of POBA-GA adversarial training and ensemble adversarial training, and prove that the defense capability of the ensemble adversarial training is weak. Defender need add the adversarial examples generate by POBA-GA into the training data set of defense model to improve the robustness of the model. The adversarial training in this paper is achieved by retraining the model after adding the adversarial examples to the training data set. The retraining data set has 20 different classes, each class consisting of 900 normal examples and 100 adversarial examples. In the ensemble adversarial training, we use FGSM, SJMA, Deep-Fool, BIM and C\&W generated 5*20 adversarial examples for each class. In the POBA-GA adversarial training, we added 100 POBA-GA adversarial examples to the training data set. Both POBA-GA adversarial training and ensemble adversarial training use a batch size of 100 and $STEPS$ (iterations number) of 10,000. The $LR$(Learning Rate) of this paper is calculated by Eq.~\ref{equ:Eq_LearnRate}.

\begin{equation}
\begin{array}{c}
LR=BLR*e^{-\frac{i*in(0.1/MLR)}{STEPS}}
\label{equ:Eq_LearnRate}
\end{array}
\end{equation}
where $BLR$ is the Base Learning Rate, $MLR$ is the Minimum Learning Rate and $i$ is the current iterations number. In this paper we make $BLR=0.1$, $MLR=0.001$. The adversarial training process is shown in Figure~\ref{fig:defence}.
}

\textbf{Ensemble adversarial training:} Generate adversarial examples to classifier the adversarial examples generated by ensemble attacks and normal examples. In our case, FGSM, SJMA, DeepFool, BIM, C\&W are adopted as ensemble model to attack $TM$.

\textbf{POBA-GA adversarial training:} Adversarial training is applied to $TM$ based on adversarial examples generated by POBA-GA. POBA-GA can provide high-quality adversarial examples for training which could improve defensibility of the target model.

Table~\ref{my-label:defensive} shows the target model's defensibility comparison after ensemble adversarial training and POBA-GA adversarial training. Attack success rate of almost all attacks are decreased significantly compared with before adversarial training (in Table~\ref{my-label:perturbation}). From the experiment result, we can also find that the attack success rate of MNIST and CIFAR-10 are lower than ImageNet64. This is mainly because they have fewer pixels and similar adversarial examples, while ImageNet64 has more pixels and can choose more ways to change. We can find that the adversarial training based on POBA-GA attacks is less effective than ensemble adversarial training on white-box attack. This is mainly because the ensemble adversarial training is implemented based on the adversarial examples generated by the white-box attack. However, the adversarial defense against black-box attacks is better than the ensemble defense. The most important thing is that adversarial defense reduce the attack success rate of POBA-GA to about 30\%.

\begin{table*}[!ht]
\centering
\caption{The defensive comparison of ensemble adversarial training and POBA-GA adversarial training.}
\label{my-label:defensive}
\begin{tabular}{ l l l c c c c c}
\hline
\hline
\multirow{2}{*}{}                                                                 \multirow{2}{*}{Defense method}       & \multirow{2}{*}{Attack category} & \multicolumn{1}{l}{\multirow{2}{*}{algorithm}} & \multirow{2}{*}{MNIST} & \multirow{2}{*}{CIFAR-10} & \multicolumn{3}{c}{ImageNet64}    \\ \cline{6-8}
                                                                                                        &     & \multicolumn{1}{c}{}                           &                        &                        & VGG19 & Resnet50 & Inc-V3 \\ \hline
\multirow{7}{*}{\multirow{2}{*}{\begin{tabular}[c]{@{}l@{}}Ensemble \\adversarial training\end{tabular}}}
&\multicolumn{1}{l}{\multirow{4}{*}{\begin{tabular}[c]{@{}l@{}}White-box\\ Adversarial attack\end{tabular}}} & FGSM                                            & 16\%                   & 15\%                   & 20\%  & 22\%      & 22\%         \\
&\multicolumn{1}{l}{}                                                                                       & DeepFool                                        & 15\%                   & 10\%                   & 10\%  & 8\%       & 10\%         \\
&\multicolumn{1}{l}{}                                                                                       & BIM                                             & 14\%                   & 14\%                   & 18\%  & 18\%      & 20\%         \\
&\multicolumn{1}{l}{}                                                                                       & C\&W                                            & 18\%                   & 18\%                   & 22\%  & 20\%      & 20\%         \\ \cline{2-8}

&\multirow{3}{*}{\begin{tabular}[c]{@{}l@{}}Black-box \\ Adversarial attack\end{tabular}}                     & Boundary                                        & 38\%                   & 36\%                   & 52\%  & 50\%      & 50\%         \\
                                                                                                            & & ZOO                                             & 64\%                   & 60\%                   & 78\%  & 80\%                                                                                          & 78\%         \\
                                                                                                            & & AutoZOOM                                             & 60\%                   & 59\%                   & 79\%  & 76\%                                                                                          & 78\%         \\
                                                                                                            & & POBA-GA                                           & \textbf{70\%}                   &\textbf{72\%}                   & \textbf{88\%}  & \textbf{86\%}      & \textbf{85\%}         \\ \hline

\multirow{7}{*}{\multirow{2}{*}{\begin{tabular}[c]{@{}l@{}}POBA-GA \\adversarial training\end{tabular}}}
&\multicolumn{1}{l}{\multirow{4}{*}{\begin{tabular}[c]{@{}l@{}}White-box\\ Adversarial attack\end{tabular}}} & FGSM                                            & 32\%                   & 35\%                   & 50\%  & 54\%      & 52\%         \\
&\multicolumn{1}{l}{}                                                                                       & DeepFool                                        & 33\%                   & 31\%                   & 36\%  & 34\%       & 38\%         \\
&\multicolumn{1}{l}{}                                                                                       & BIM                                             & 35\%                   & 33\%                   & 42\%  & 44\%      & 40\%         \\
&\multicolumn{1}{l}{}                                                                                       & C\&W                                            & 32\%                   & 35\%                   & 52\%  & 50\%      & 51\%         \\ \cline{2-8}

&\multirow{3}{*}{\begin{tabular}[c]{@{}l@{}}Black-box \\ Adversarial attack\end{tabular}}                     & Boundary                                        & 37\%                   & 38\%                   & 51\%  & 52\%      & 54\%         \\
                                                                                                             && ZOO                                             & 44\%                   & 42\%                   & 74\%  & 74\%      & 72\%         \\
                                                                                                            & & AutoZOOM                                             & 46\%                   & 40\%                   & 76\%  & 73\%                                                                                          & 76\%         \\
                                                                                                             && POBA-GA                                           & \textbf{26\%}                 & \textbf{28\%}                 & \textbf{34\%} & \textbf{32\%}      & \textbf{32\%}         \\ \hline
\hline
\end{tabular}
\end{table*}

\subsection{Facial recognition application}
{Our experiments were mainly carried out on data sets such as VGG19, inc-V3 and Resnet50. However, such detectors are not widely used in practical applications. Therefore, we expand the application scenario of POBA-GA.} In recent years, facial recognition-based identity authentication systems have become popular and widely used in life~\cite{deng2017fine, zhou2018age}, so the security of face recognition systems has become increasingly important.

\begin{figure}[H]
\centering
  \includegraphics[width=0.8\linewidth]{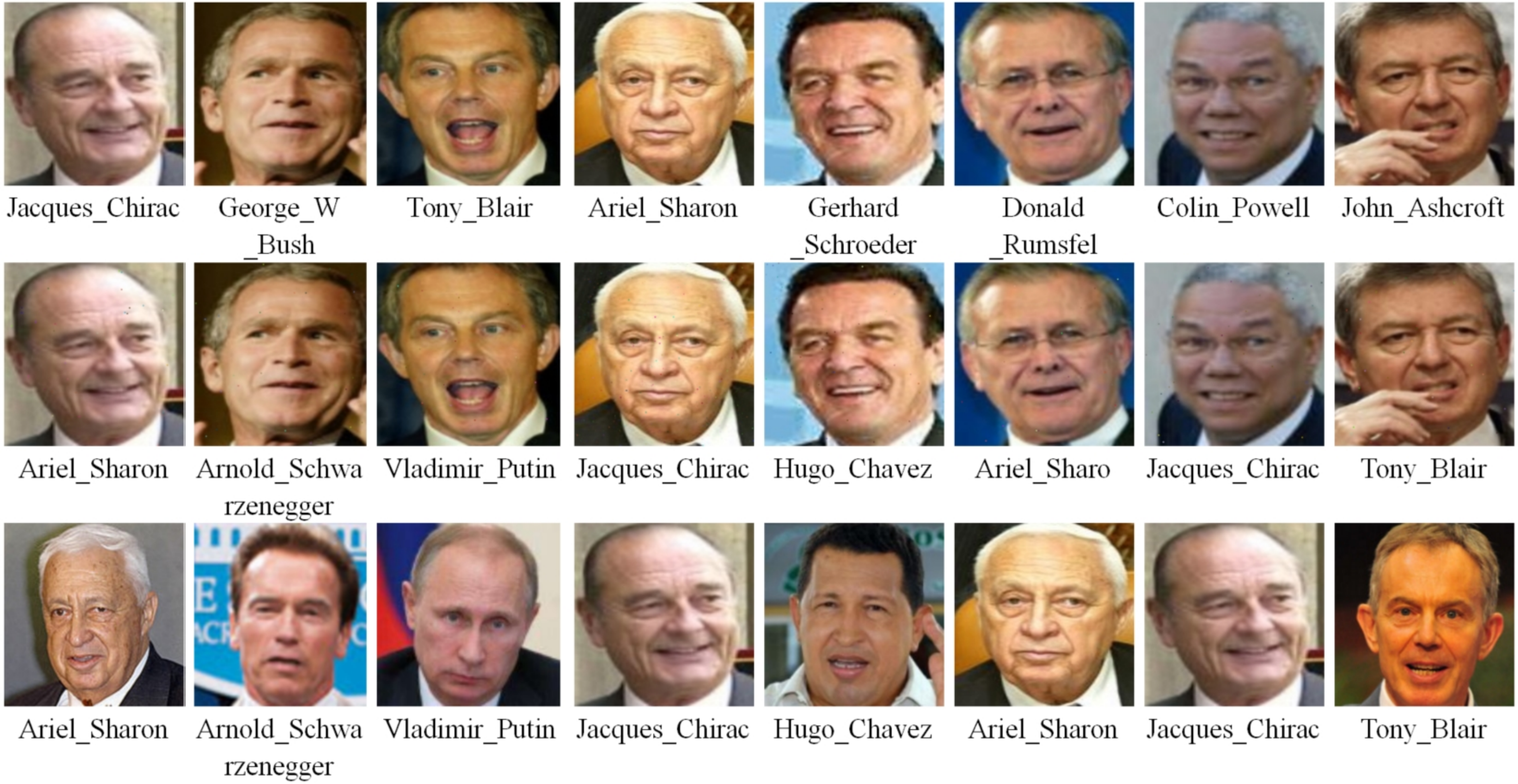}
\caption{Facial recognition application. The photos of first row are the original images, the photos of second row are adversarial examples, and the photos of third row are the faces corresponding to the wrong label.}
\label{fig:Application}
\end{figure}

{The experiment selected the Wild dataset (LFW) as the experimental data set.}The Labeled Faces in the Wild dataset (LFW)~\cite{huang2008labeled} containing more than 5,000 faces and more than 10,000 images. We use POBA-GA to attack LFW faces and generate corresponding adversarial examples. Figure~\ref{fig:Application} shows the original image and their true labels, and the adversarial examples generated by POBA-GA and the target model predicted label. From the figure we can see that POBA-GA can indeed achieve face attacks through perturbation optimization, which has strong applicability.

\section{Conclusion\label{Conclusion}}
Adversarial attack against deep model can cause fundamental errors. We focus on black-box attacks since it is more operable and harder to defend. Different from current black-box adversarial attacks, we propose a novel evolutionary algorithm based adversarial example generation method for black-box attack implementation. A perturbation optimized black-box attacks (POBA-GA) is put forward against deep neural networks. Abundant experiments are carried out compared with classic white-box and black-box adversarial attack methods. The results prove that POBA-GA has higher attack success rate than other attack methods. It can achieve 100\% attack success rate in CIFAR-10 and MNIST classification models, and it can achieve 96\% attack success rate on ImageNet64 black-box method. In both attack success rate and perturbation control, POBA-GA has better performance than existing black-box attack. For further study, we will study on POBA-GA's attack transferability on different target models.

\section*{Acknowledgment}
This work is partially supported by National Natural Science Foundation of China (61502423, 61572439), Zhejiang Natural Science Foundation(LY19F020025), Signal Recognition Based on GAN, Deep Learning for Enhancement Recognition Project, Zhejiang University Open Fund(2018KFJJ07), and Zhejiang Science and Technology Plan Project (2017C33149).

\section*{References}

\bibliography{mybib}

{
\section*{Appendix}
\subsection*{Parameter adjustment}
Figure~\ref{fig:p_c} shows the effect of $P_c$ on experimental results. Figure~\ref{fig:p_c} (a) shows the best of the fitness function of the parent adversarial the example. Figures~\ref{fig:p_c} (b) and (c) show the perturbation and attack performance $P$ of examples with the best fitness function value.

\captionsetup[figure]{singlelinecheck=off}
\begin{figure}[H]
\centering
\captionsetup{justification=centering}
  \subfloat[Best fitness function value]{\includegraphics[width=0.48\linewidth]{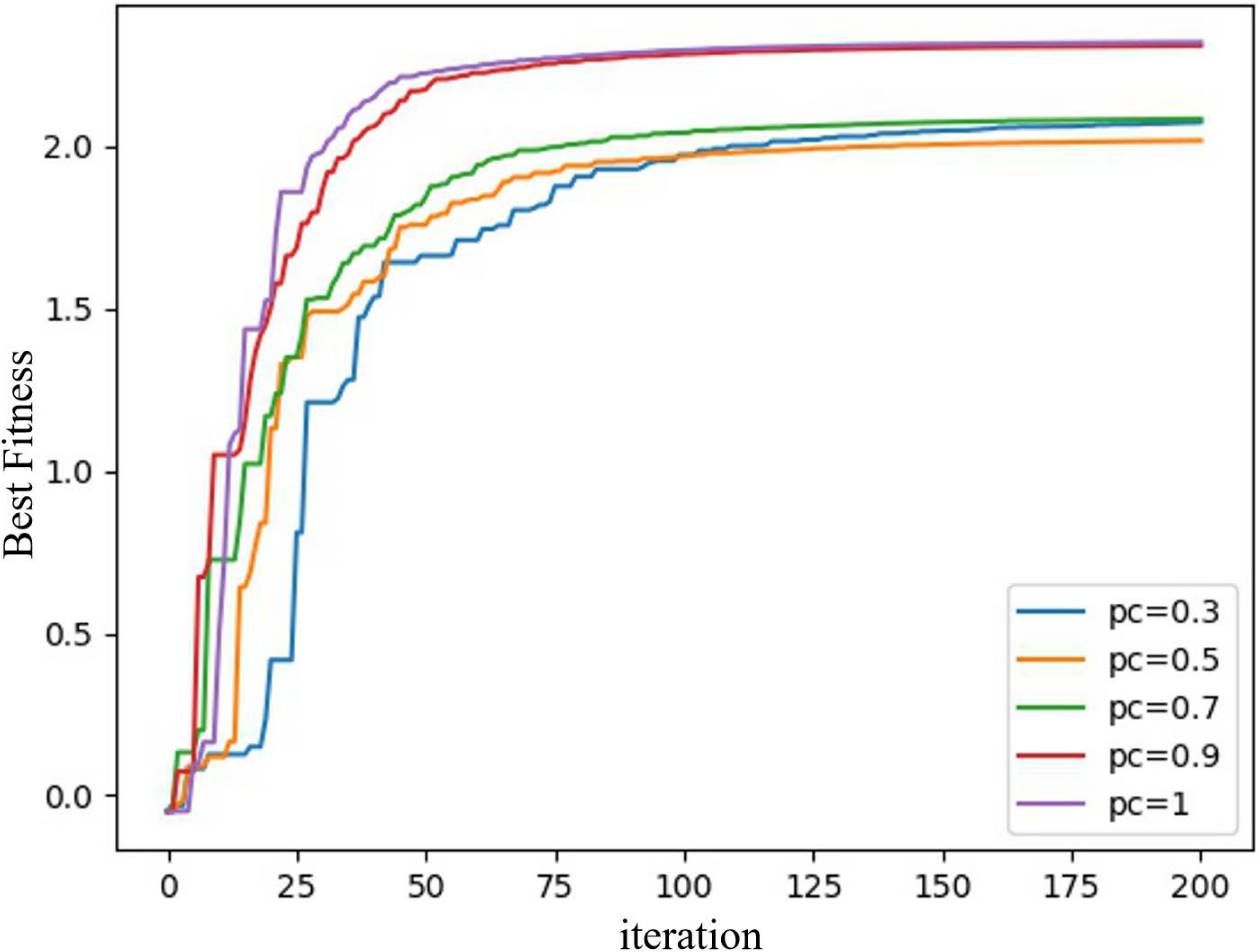}}\vspace{1ex}
  \subfloat[Perturbation]{\includegraphics[width=0.48\linewidth]{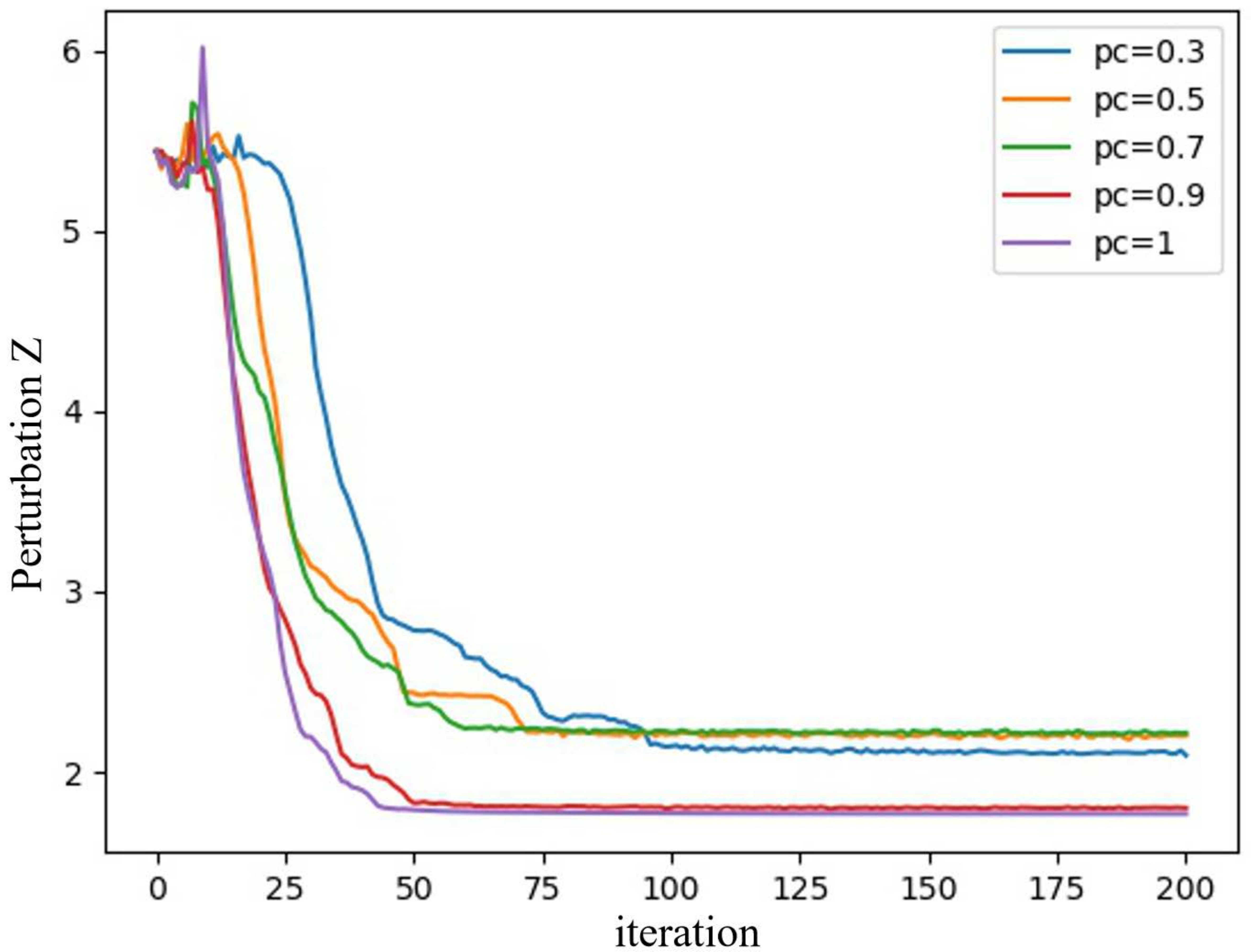}}\vspace{1ex}
  \subfloat[Attack performance $P$]{\includegraphics[width=0.48\linewidth]{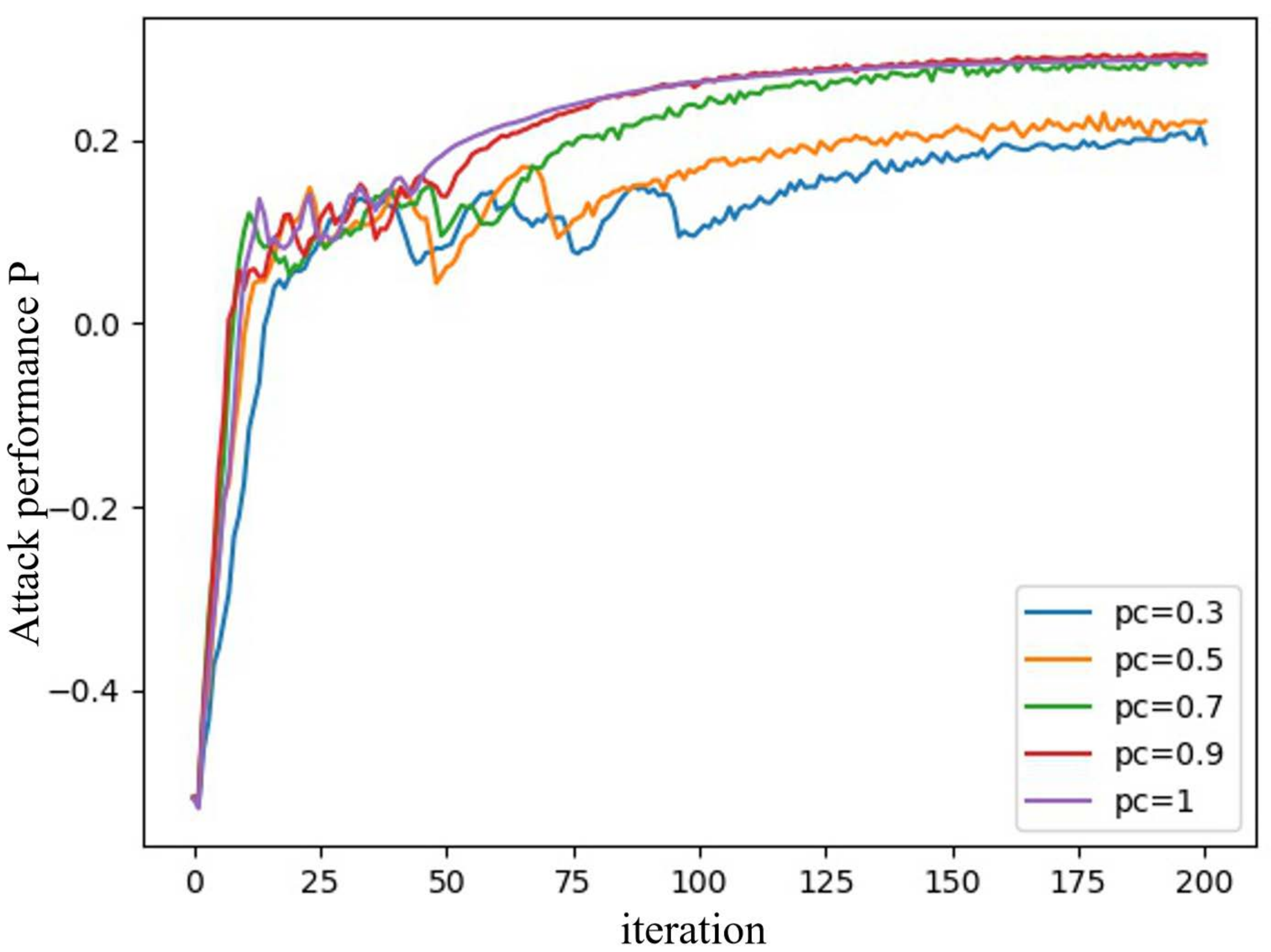}}\vspace{1ex}
\caption{The effect of $P_c$ on experimental results.}
\label{fig:p_c}
\end{figure}

}

\end{document}